\documentclass[english,floatfix,twocolumn,showpacs, aps, prx, reprint, superscriptaddress, longbibliography]{revtex4-2}
\usepackage{amsmath}
\usepackage{hyperref}
\hypersetup{colorlinks=true, citecolor=blue, urlcolor=blue, linkcolor=blue, breaklinks=true}
\usepackage{amssymb}
\usepackage{graphicx}
\newcommand{\commentout}[1]{}
\usepackage{booktabs, footnote}
\usepackage{physics}

\begin{document}
\title{Large-scale quantum reservoir learning with an analog quantum computer}
\author{Milan Kornjača$^\dagger$}
\thanks{These authors contributed equally.\\
$^\dagger$Email address:
\href{mailto:swang@quera.com}{swang@quera.com};
\href{mailto:fliu@quera.com}{fliu@quera.com};
\href{mailto:mkornjaca@quera.com}{mkornjaca@quera.com}
}
\affiliation{QuEra Computing Inc., 1284 Soldiers Field Road, Boston, MA, 02135, USA}
\author{Hong-Ye Hu}
\thanks{These authors contributed equally.\\
$^\dagger$Email address:
\href{mailto:swang@quera.com}{swang@quera.com};
\href{mailto:fliu@quera.com}{fliu@quera.com};
\href{mailto:mkornjaca@quera.com}{mkornjaca@quera.com}
}
\affiliation{QuEra Computing Inc., 1284 Soldiers Field Road, Boston, MA, 02135, USA}
\affiliation{Department of Physics, Harvard University, Cambridge, MA 02138, USA}
\author{Chen Zhao}
\author{Jonathan Wurtz}
\author{Phillip Weinberg}
\author{Majd Hamdan}
\author{Andrii Zhdanov}
\author{Sergio  H. Cantu}
\author{Hengyun Zhou}
\affiliation{QuEra Computing Inc., 1284 Soldiers Field Road, Boston, MA, 02135, USA}
\author{Rodrigo Araiza Bravo}
\affiliation{Department of Physics, Harvard University, Cambridge, MA 02138, USA}
\author{Kevin Bagnall}
\author{James I. Basham}
\author{Joseph Campo}
\author{Adam Choukri}
\author{Robert DeAngelo}
\author{Paige Frederick}
\author{David Haines}
\author{Julian Hammett}
\author{Ning Hsu}
\author{Ming-Guang Hu}
\author{Florian Huber}
\author{Paul Niklas Jepsen}
\author{Ningyuan Jia}
\author{Thomas Karolyshyn}
\author{Minho Kwon}
\author{John Long}
\author{Jonathan Lopatin}
\author{Alexander Lukin}
\author{Tommaso Macrì}
\author{Ognjen Markovi\'c}
\author{Luis A. Martínez-Martínez}
\author{Xianmei Meng}
\author{Evgeny Ostroumov}
\author{David Paquette}
\author{John Robinson}
\author{Pedro Sales Rodriguez}
\author{Anshuman Singh}
\author{Nandan Sinha}
\author{Henry Thoreen}
\author{Noel Wan}
\author{Daniel Waxman-Lenz}
\author{Tak Wong}
\author{Kai-Hsin Wu}
\author{Pedro L. S. Lopes}
\author{Yuval Boger}
\author{Nathan Gemelke}
\author{Takuya Kitagawa}
\author{Alexander Keesling}
\affiliation{QuEra Computing Inc., 1284 Soldiers Field Road, Boston, MA, 02135, USA}
\author{Xun Gao}
\affiliation{JILA and Department of Physics, University of Colorado, Boulder, Colorado 80309, USA}
\author{Alexei Bylinskii}
\affiliation{QuEra Computing Inc., 1284 Soldiers Field Road, Boston, MA, 02135, USA}
\author{Susanne F. Yelin}
\affiliation{Department of Physics, Harvard University, Cambridge, MA 02138, USA}
\author{Fangli Liu$^\dagger$}
\affiliation{QuEra Computing Inc., 1284 Soldiers Field Road, Boston, MA, 02135, USA}
\author{Sheng-Tao Wang$^\dagger$}
\affiliation{QuEra Computing Inc., 1284 Soldiers Field Road, Boston, MA, 02135, USA}
\date{\today}


\begin{abstract}

Quantum machine learning has gained considerable attention as quantum technology advances, presenting a promising approach for efficiently learning complex data patterns. Despite this promise, most contemporary quantum methods require significant resources for variational parameter optimization and face issues with vanishing gradients, leading to experiments that are either limited in scale or lack potential for quantum advantage. To address this, we develop a general-purpose, gradient-free, and scalable quantum reservoir learning algorithm that harnesses the quantum dynamics of neutral-atom analog quantum computers to process data. We experimentally implement the algorithm, achieving competitive performance across various categories of machine learning tasks, including binary and multi-class classification, as well as timeseries prediction. Effective and improving learning is observed with increasing system sizes of up to 108 qubits, demonstrating the largest quantum machine learning experiment to date. We further observe comparative quantum kernel advantage in learning tasks by constructing synthetic datasets based on the geometric differences between generated quantum and classical data kernels. Our findings demonstrate the potential of utilizing classically intractable quantum correlations for effective machine learning. We expect these results to stimulate further extensions to different quantum hardware and machine learning paradigms, including early fault-tolerant hardware and generative machine learning tasks.

\end{abstract}

\maketitle

\section*{Introduction}
With the increasing power of machine learning and the emergence of quantum computers, there is growing interest in exploring the intersection of these fields through quantum machine learning (QML) \cite{Biamonte2017}. The main prospect for QML lies in the potential of even near-term quantum devices \cite{Preskill_NISQ_2018} to produce correlations that would otherwise require exponential classical compute~\cite{Arute2019, Huang2021b, Liu2021, Anschuetz2023,Huang2021}. With the notable exception of explicitly quantum tasks \cite{Huang2022}, however, this potential remains mostly untapped by experiments. Currently, the predominant QML algorithms designed for noisy quantum hardware are variational algorithms that leverage parametrized circuits within a hybrid quantum-classical framework~\cite{Cerezo2021}. These approaches, however, face several significant obstacles. Fundamentally, issues such as noise, entanglement-induced barren plateaus, and complex training landscapes present considerable challenges to the trainability of these models~\cite{McClean2018, Ortiz2021, Anschuetz2022, Larocca2024}. Practically, the necessity to estimate gradients puts a strain on the already limited resources of near-term quantum hardware, leading to demonstrations that are often limited in both scale and performance~\cite{Havlicek2019, Shen2024}.

\begin{figure*}[htb]
\centering
\includegraphics[width=\textwidth]{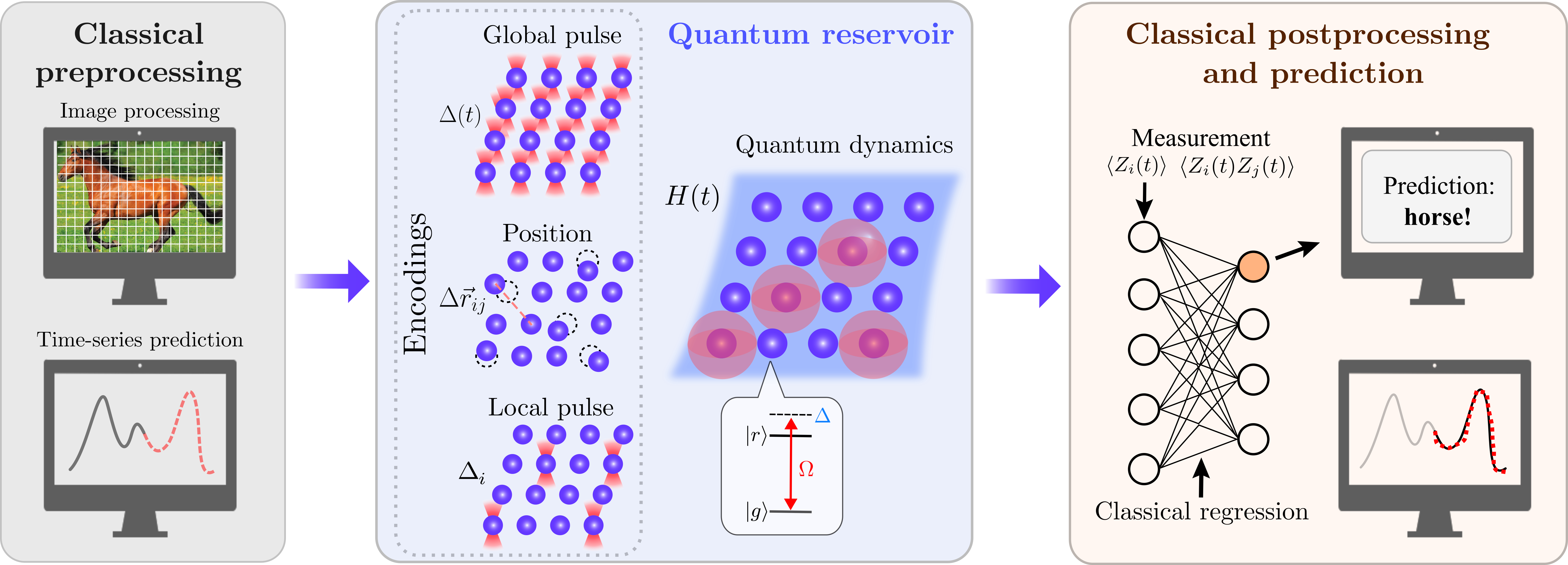}
\caption{\textbf{Overview of the quantum reservoir computing (QRC) algorithm with neutral atoms.} The QRC algorithm pipeline contains three steps -- classical preprocessing (left), quantum reservoir (center), and classical postprocessing and prediction (right). In the preprocessing step, data features are brought into a form readily encoded to the neutral-atom analog quantum computer. They may require optional dimensional reduction for high-dimensional data (such as images, top) or feature engineering and selection (such as data windowing for timeseries, bottom). The encoding of the data features proceeds by three methods, i.e, encoding into the time profile of the global detuning pulse, the interaction strengths by atom position modulation, and the local pattern of the detuning pulse. The quantum system serving as the reservoir is then evolved over varying time periods and probed through repeated projective measurements. In the third step, the measurement outputs are processed classically to provide expectation values of local observables that form a set of QRC embeddings. The embeddings are subsequently used as inputs to a simple and fast classical training step, for which we typically employ linear support vector machines or regression. 
The trained models are tested and used for inference by processing additional data through the QRC pipeline and evaluating classical outputs based on obtained embeddings.
}
\label{fig:Fig1_QRC}
\end{figure*}

Classical reservoir computing (RC) represents an alternative machine learning paradigm that circumvents the need for costly gradient optimization~\cite{Jaeger2004}\commentout{\cite{Jaeger2004, Jaeger2007, Lukosevicius2009, Tanaka2019, Gauthier2021}}. It processes data by employing a complex dynamical system known as the \emph{reservoir} to transform input data nonlinearly and expand the embedding space dimension. This transformation enhances the features used in training, typically conducted via simple linear regression. RC methods are primarily used for predicting chaotic systems and forecasting time series~\cite{Jaeger2001, Arcomano2023}, yet they can also tackle general tasks such as pattern recognition and generation~\cite{Jalalvand2015, Schaetti2016}. As this framework can mitigate many of the aforementioned challenges in QML, quantum reservoir computing (QRC) has recently emerged to be a promising approach suitable for near-term quantum hardware~\cite{Fujii2017, Martinez2021,  Bravo2022,  Dudas2023, Mujal2023, Hu2023, Senanian2023, Yasuda2023}. In QRC, a complex quantum system replaces the classical reservoir, exploiting dynamics in the exponentially large Hilbert space to produce classically intractable correlations at the outputs. These prior studies, however, are limited to either numerical simulations or small-scale experiments in the classically tractable regime with unclear prospects for quantum utility. 

Here, we explore a QRC paradigm based on the complex quantum dynamics of a neutral-atom analog quantum computer. Several seminal results have been demonstrated on such devices, showing their unique capabilities in quantum simulation and optimization~\cite{Scholl2021, Ebadi2021, Semeghini2021, Ebadi2022}. In the case of QML applications studied in this work, we co-design the data encodings and physical parameters in the neutral-atom Hamiltonian. We report three major advances. First, we demonstrate effective learning in our experiments with up to 108 qubits, representing a substantial leap over previously reported QML results~\cite{ Johri2021, Huang2022, Hu2023, Senanian2023,Yasuda2023,Albrecht2023, Haug2023, Jennifer24}. Second, we experimentally observe comparative quantum kernel advantage by comparing the QRC-generated kernel with classical kernels using a recently proposed synthetic dataset construction procedure~\cite{Huang2021}. This shows the existence of datasets for which non-classical correlations of QRC can be utilized for effective machine learning even on current, noisy quantum hardware. Lastly, we show the versatility and noise-resilience of our QRC framework in experiments over a diverse set of tasks and data. A universal parameter regime, informed by physical insights, eliminates the need for any parameter optimization in the quantum part, resulting in substantial savings of quantum resources.

\section*{Co-designed Algorithm}

Our QRC algorithm is employed for supervised machine learning tasks. All experiments are performed on a publicly accessible analog quantum computer, Aquila~\cite{Wurtz2023}. The framework is presented schematically in Fig.~\ref{fig:Fig1_QRC}. The first step is preprocessing the data into a form suitable for supervised tasks. The training and testing datasets consist of pairs of data, $\{(x_i[n],y_j[n])\} $, where $x_i[n]$ are input feature vectors, and $y_j[n]$ are the corresponding labels; $n$ enumerates the data, while $i \, (j)$ indexes feature (label) vector components. For example, in image classification problems, the input data $x_i$ are the image pixel values, and $y_j$ label image objects (classes). Depending on the data type, feature dimension reduction may be needed to fit the problem to hardware-tractable sizes.

\begin{figure*}[htb]
\centering
\includegraphics[width=1.0\textwidth]{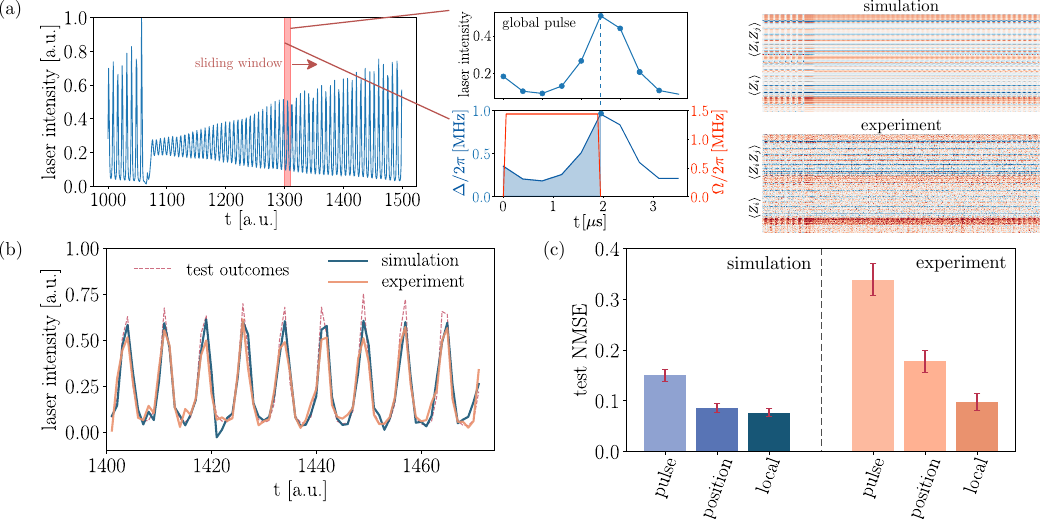}
\caption{\textbf{Timeseries prediction with QRC and encoding performance comparison.} (a) The pipeline for global pulse encoding timeseries prediction with QRC. The dataset chosen is a part of the Santa Fe laser timeseries \cite{Weigend1993-bg}, with feature vectors constructed by sliding windows and the task set to one-step prediction. The profile of the window feature is encoded into the piecewise linear global detuning pulse. The select local observables obtained by probing the quantum evolution are shown for exact simulation and experiment, with the vertical axis corresponding to different embedding components and the horizontal to the timeseries steps. (b) An example of test outcomes predicted by local pulse encoded QRC with 12 qubits, compared to the equivalent finite-sampled simulation (110 shots per datapoint) and the true outcomes. (c) Comparison of the normalized mean-square error (NMSE, lower is better) for three different QRC encodings in finite-sampled simulation and experiment.
}
\label{fig:Fig2_timeseries}
\end{figure*}

The neutral-atom QRC proceeds by encoding the data features into the parameters of the Rydberg Hamiltonian, given by \cite{Wurtz2023}
\begin{align}\label{eq:RydHam}
    H(t)&=\dfrac{\Omega(t)}{2}\sum_j \left(\ket{g_j}\bra{r_j}+\ket{r_j}\bra{g_j}\right)\cr
    &+\sum_{j<k}V_{jk}n_jn_k-\sum_j \left[\Delta_{\mathrm{g}}(t) + \alpha_j\Delta_{\mathrm{l}}(t)\right] n_j,
\end{align}
where $\Omega(t)$ is the global Rabi drive amplitude between a ground ($\ket{g_j}$, with $j$ indexing atoms) and a highly-excited Rydberg state of an atom ($\ket{r_j}$), $n_j=\ket{r_j}\bra{r_j}$, while  $V_{jk}=C/\lVert\mathbf{r}_j-\mathbf{r}_k\rVert^6$ describes the van der Waals interactions between atoms. The detuning is split into the global term, $\Delta_{\mathrm{g}}(t)$ and the site-dependent term $\Delta_{\mathrm{l}}(t)$, with site modulation $\alpha_j\in[0,1]$.  The system's tunable control parameters allow us to explore three distinct encoding schemes (see Fig.~\ref{fig:Fig1_QRC}): 

(1) \emph{Global pulse encoding} that is typically implemented by mapping data features into the time-varying profile of a global detuning pulse, $\Delta_{\mathrm{g}}(t_i)=\Delta_{\mathrm{g}}^{\mathrm{max}} x_i$. With features encoded as pulse parameters at different times, the encoding capacity is independent of the system size ($N_q$) and depends on the pulse protocol.

(2) \emph{Position encoding} that proceeds by engineering atom positions such that the Rydberg interaction strength is modulated according to the data features. Nearest-neighbor interaction strengths are modulated with $V_{il}=V^{(0)}(1+\lambda x_i)$, where $V^{(0)}$ represents the bare (unmodulated) interaction strength,  $l$ denotes one nearest neighbour of $i$, and $\lambda$ is an encoding scale. Nearest neighbor interactions in a one-dimensional $N_q$-qubit system allow the encoding of $N_q-1$ features, while our two-dimensional implementations typically employ a sequence of one-dimensional row encodings (see Supplementary Information).

(3) \emph{Local pulse encoding} that is implemented through site-dependent local detunings with $\alpha_i\Delta_{\mathrm{l}}(t)=\Delta_{\mathrm{l}}^{\mathrm{max}} x_i$; an $N_q$-qubit system can thus encode $N_q$ features.

After data encoding, the quantum system evolves from an all-ground state with the correspondingly designed Rydberg Hamiltonian. The quantum dynamics is probed in several successive timesteps, with each instance repeated in $N_s$  measurement shots. This quantum reservoir provides a platform for potentially classically intractable dynamics that induces non-linear data transformations. The measurement results are then used to obtain expectation values of local observables, typically one- and two-point correlators in the computational basis accessible directly from the measurement data -- $\langle Z_j \rangle$ and $\langle Z_j Z_k\rangle$, where $Z_j=2n_j-I_j$. The local observables then form the data-embedding vectors, {$u_i[n]$}, where $i$ enumerates different correlators and probe times. The quantum reservoir step is closely connected to the notion of a QML kernel \cite{Liu2021,Huang2021,Jennifer24}. In fact, the reservoir embeddings define a kernel matrix on the data with $K(\mathbf{x}[n],\mathbf{x}[m])=\langle \mathbf{u}[n], \mathbf{u}[m] \rangle$, where $\langle \cdot,\cdot\rangle$ is the inner product in the embedding vector space. The kernel matrix redefines the notion of distance and geometry between data. Kernel geometry has been central in a recent theoretical proposal that establishes early quantum advantage prospects \cite{Huang2021}, which we apply in an experimental QRC setting.

The final step of the QRC algorithm is classical postprocessing, consisting of training and inference on the data embeddings, requiring no optimization loops on quantum hardware. In our scheme, this typically entails using simple and readily trainable linear models, such as linear support vector machines (SVMs) \cite{Boser1992, Libsvm2011}, with task-dependent details described in the Supplementary Information. More generally, any classical supervised machine learning algorithm can be used. The trained model is then used for inference on the test set quantum embeddings and evaluating QRC performance. 

\section*{Results on quantum hardware}
\label{sec:resultshardware}

We now proceed to comprehensively demonstrate the viability of the QRC approach and determine its performance limits and utility potential. We perform extensive studies on the neutral-atom quantum hardware, together with accompanying numerical simulations. Our first goal is to determine the parameter regime of the Hamiltonian (Eq.~\ref{eq:RydHam}) in which the QRC algorithm performance is optimal and robust to noise. To this end, we perform extensive numerical simulations, with the main result being the discovery of a universal parameter regime, as described in the Supplementary Information. This regime is manifested as a wide island in the parameter space where QRC performance is optimal across multiple datasets we consider. Its defining feature is the three energy scales being of comparable magnitude -- the quantum state mixing scale ($\Omega$), the entangling scale  ($V$), and the encoding scale. Furthermore, we find that within such an optimality island, further fine-tuning of parameters leads to negligible gains. The physical guidance enables us to establish QRC as a quantum gradient-free approach and avoid costly hyperparameter training on hardware.

\begin{figure*}[htb]
\centering
\includegraphics[width=1.0\textwidth]{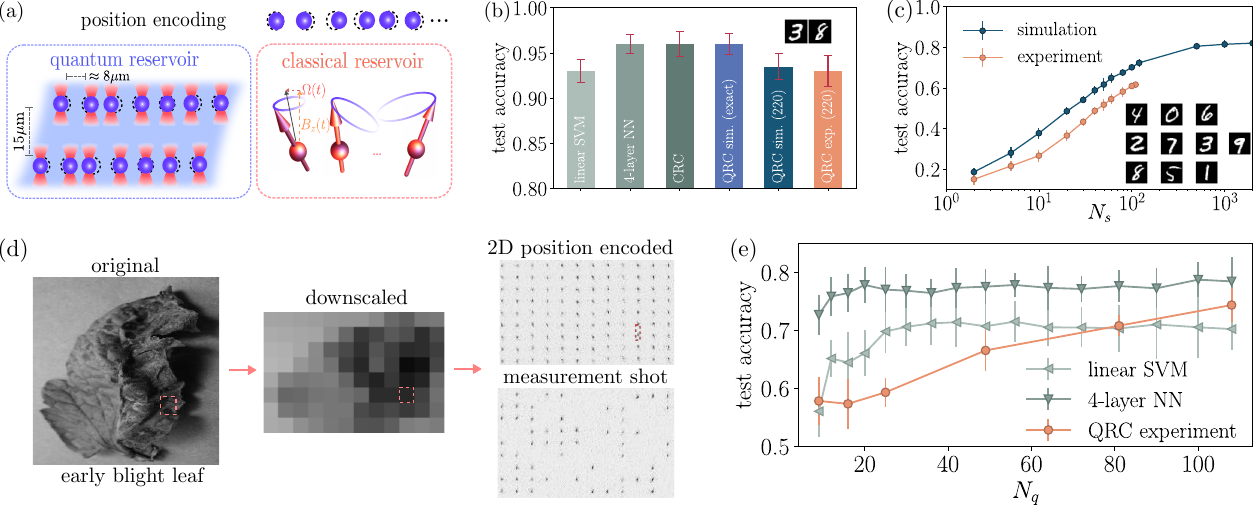}
\caption{\textbf{Image classification with QRC and qubit number scaling.} (a) The MNIST~\cite{Deng2012} images of handwritten digits are downsampled to feature vectors that are encoded into the modulation of the nearest neighbor Rydberg interaction strengths via the position encoding. The quantum reservoir consists of parallel, well-separated neutral-atom chains evolving under the Rydberg Hamiltonian. The equivalent classical spin reservoir (CRC), where the vector spins precess in the external and neighbor magnetic field, is simulated for comparison. (b) The test classification accuracy of several classical machine learning methods and QRC on the 3/8-MNIST binary classification tasks. (c) The test accuracy of QRC simulation and experiment as a function of the number of shots drawn per datapoint, $N_s$, for the 10-class MNIST classification task.  (d) The QRC performance scaling with the number of qubits was probed on the tomato disease task, with three classes selected from the plant village dataset \cite{hughes2016tomato}. The data features, the pixel values of the downscaled images of different sizes, were encoded in the vertical interaction strengths of a 2D atom array. An example experimental image of the position-encoded atom array is shown, together with one measurement shot after the quantum reservoir dynamics. (e) The test accuracy as a function of the qubit number, $N_q$, or the equivalent feature dimension, for the QRC performed on the experiment, linear SVM, and 4-layer feedforward neural network.
}
\label{fig:Fig3_classification}
\end{figure*}

\subsection*{Timeseries prediction and encoding comparison}

Classical reservoir computing has been successfully applied to tasks such as time series analysis and complex system prediction~\cite{Antonik2018, Nadiga2021, Viehweg2023, Arcomano2023}. Motivated by this and as a proof-of-concept, we apply QRC to process the Santa Fe timeseries task \cite{Weigend1993-bg}, which represents the intensity profile of a laser in a chaotic regime. The portion of the task we process experimentally is shown in Fig.~\ref{fig:Fig2_timeseries}(a).  We apply and compare three different QRC encodings on this task: the global pulse, position, and local pulse encodings.

Processing of the time series in all three cases starts with feature extraction from a time window with $d$ time-steps and specifying a future 1-step point as the prediction target. The global pulse provides a natural representation of the time series by encoding the time-window features directly into the piece-wise linear global detuning waveform, as shown in Fig.~\ref{fig:Fig2_timeseries}(a). The resulting embeddings offer an intuitive insight into the QRC method [see Fig.~\ref{fig:Fig2_timeseries}(a), right]. The numerically-simulated embeddings clearly express features of the timeseries, with each local observable providing a distinct ``look" on the same data.  The success of QRC is derived from this ability to transform data non-linearly, representing its different aspects. The hardware-generated embeddings show excellent agreement with the exact simulation, with some additional noise due to finite sampling and experimental noise.

An example of QRC predictions generated on the test dataset is shown in Fig.~\ref{fig:Fig2_timeseries}(b) for the local detuning encoding experiment and finite-sampled simulation. QRC effectively captures the features of the time series, although the main outliers are concentrated near the end of its dynamic range, where the effect of sampling noise is most pronounced. The comparison of the simulated and experimental performance for all three encodings is presented in Fig.~\ref{fig:Fig2_timeseries}(c). The relative order of encoding performance is preserved between the simulation and experiment, with the experiment showing additional performance decay for global pulse encoding and, to a lesser extent, position encoding.

The performance across different encodings reflects general considerations of experimental noise applicable to various QRC tasks. In simulations, both exact and finitely sampled cases show approximate equivalence between position and local encodings, with global pulse encoding performing less effectively. This discrepancy arises because the position and local encodings are applied entirely at the protocol's outset, whereas global pulse encoding gradually expresses feature vectors during evolution. Generically, quantum dynamics tends to thermalize local observables (embeddings) in the long-time limit~\cite{Deutsch_2018}. Consequently, the dynamics at later times become largely independent of the pulse shape, causing significant portions of the feature vector to be encoded in a lossy manner. In the experiment, decoherence leads to even more lossy encoding of the later-time input features, acting as an additional systematic noise source for the global pulse encoding. The difference between the position and local pulse encodings observed in experiments is more subtle and results from the shot-to-shot atom position fluctuations~\cite{Wurtz2023}. Although these fluctuations are present for all encodings as a random noise source (mitigated by ensemble averaging), they impact the position encoding the most. Numerical simulations that model both the position fluctuations and other relevant sources of coherent noise confirm this picture and qualitatively reproduce the experimental performance (see Supplementary Information). The overall good agreement between the experiment and simulation highlights the robustness of QRC under random noise sources. The protocol primarily exhibits sensitivity to noise channels that produce inconsistent or lossy embeddings. In the Supplementary Information, we introduce a directly calculable measure, based on a statistical correlation between embeddings, that can help characterize and reduce the noise related to hardware consistency.

\subsection*{Image classification and qubit scaling performance}

We demonstrate the versatility of our QRC algorithm by applying it to various image classification tasks. These include binary and 10-class classifications of the MNIST handwritten digits dataset~\cite{Deng2012} [Fig.~\ref{fig:Fig3_classification}~(a)-(c)] and 3-class classification of the tomato leaf disease dataset~\cite{hughes2016tomato} [Fig.~\ref{fig:Fig3_classification}~(d)-(e)]. For the MNIST dataset, image features are first downsampled (see Supplementary Information) and then position-encoded in a 9-atom qubit chain. A significant reduction in hardware runtime is achieved by performing quantum dynamics in parallel across six effectively decoupled chains, thus collecting $N_s$ measurements per embedding.

We evaluate QRC's performance by comparing it to several classical methods: a linear SVM baseline, a four-layer feedforward neural network, and the equivalent classical spin reservoir (CRC). The CRC, as shown in Fig.~\ref{fig:Fig3_classification}(a), is derived from the QRC by treating the spins as classical vectors, effectively taking the infinite spin limit; classical spins precess in the instantaneous magnetic field determined by the Rabi frequency, detuning, and the interactions. The CRC pipeline is otherwise equivalent to the one described for QRC. All states of the CRC have a one-to-one mapping to the manifold of all product states, making the QRC-CRC comparison a heuristic measure of the importance of quantum entanglement.

The experimental results on the MNIST dataset are presented in Fig.~\ref{fig:Fig3_classification}(b)-(c). Initial probing focuses on a binary classification subtask within the MNIST dataset, the classification of digits 3/8, which serves as a confirmation of the QRC pipeline. Results in Fig.~\ref{fig:Fig3_classification}(b) show a clear separation between the performance of linear and non-linear methods. The QRC method, when simulated exactly, achieves the performance benchmarks set by classical methods for the task. However, the practical performance is constrained by the finite number of measurement shots. With $N_s=220$, the experiment achieves a test accuracy of 0.935, which is within half a percent of the performance achieved in simulations under the same $N_s$. This successful result in the 3/8 classification task demonstrates the robustness of the QRC, even in the presence of significant experimental noise.

To further investigate the impact of the finite number of measurements on algorithm performance, we explore how QRC performance scales with the number of measurements in both experiments and simulations for the 10-class MNIST task, as depicted in Fig.~\ref{fig:Fig3_classification}(c). The task, being considerably more complex than the 3/8 binary classification, serves as a more stringent test of both experimental noise resilience and the measurement requirements necessary for good performance. The data shows that the QRC performance tends to plateau at approximately $N_s \sim 1000$, a trend consistent across all datasets examined. In our experiment, we opt for $N_s\sim 100$ shots per data point, striking a practical balance between runtime and performance. The observed performance disparity is attributable to factors such as hardware inconsistency (see Supplementary Information) and increased sampling requirements stemming from random noise sources such as position fluctuations. Despite the extensive literature on QML algorithms, multi-class classification remains challenging on quantum hardware \cite{Pan23, Haug2023, Shen2024}. However, with the proposed QRC algorithm, we demonstrate the best-known performance in 10-class MNIST task on hardware.

Aiming to scale up QRC to large systems and evaluate its effectiveness, we consider a dataset of tomato disease classification from leaf images~\cite{hughes2016tomato}, which offers a mix of local and global disease features, as shown in Fig.~\ref{fig:Fig3_classification}(d). The features are extracted by downscaling an image to a $R_x \times R_y$ resolution and the pixels are subsequently encoded in a 2D neutral-atom system of $N_q=(R_x+1) \times R_y$ qubits via position encoding. To explore how the system scales, $N_q$ is varied from 9 to 81 in a square aspect ratio, supplemented by an experimental configuration with a $9 \times 12$ rectangular geometry.

The test accuracy in our QRC experiment, as a function of the system size, is presented in Fig.~\ref{fig:Fig3_classification}(e), where it is compared with classical linear and non-linear methods. Notably, experimental QRC performance increases with the system size up to the largest size explored. The successful implementation in a 108-qubit experiment marks a substantial increase in the QML scale~\cite{Huang2022, Hu2023,Senanian2023,Yasuda2023,Albrecht2023, Jennifer24}. It is also important to highlight that all system sizes achieve non-trivial performance, surpassing the expected 0.33 accuracy of random guessing. Moreover, the two largest system sizes not only exceed the performance of the linear SVM baseline but also approach the benchmark set by the 4-layer neural network, which has $\sim 20000$ hidden parameters.

No classical simulations are known to be able to readily simulate the quantum dynamics at the scale of our experiments, and thus, QRC hyperparameter choice is guided directly by the universal parameter regime. While the universality is validated on small ($<16$ qubits) 2D QRC systems (see Supplementary Information), the consistently improving performance with increasing system sizes testifies to the practicality of employing QRC in a truly quantum-gradient-free fashion. 

\subsection*{Comparative quantum kernel advantage}

\begin{figure}[htb]
\centering
\includegraphics[width=1.0\columnwidth]{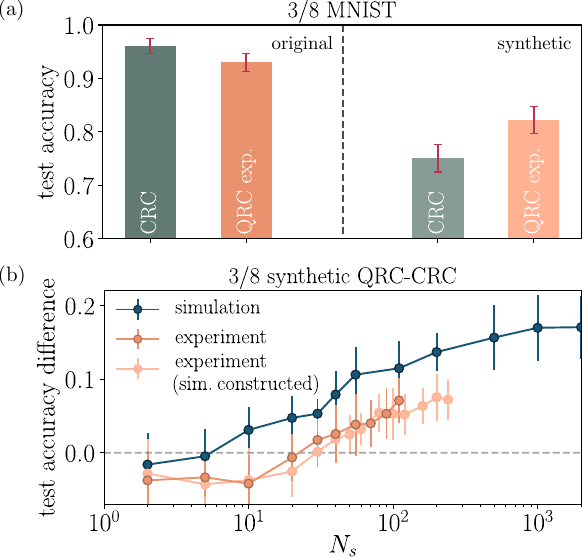}
\caption{\textbf{Comparative quantum kernel advantage with QRC.} (a) The test classification accuracy of the QRC and CRC methods on the original 3/8 binary MNIST classification and the synthetic task constructed with kernel geometry. (b) Test accuracy difference between QRC and CRC on the synthetic 3/8-binary task as a function of the number of measurements drawn per datapoint for simulation, experiment with synthetic data constructed from experimental data, and experiment with synthetic data constructed from simulation.
}
\label{fig:Fig4_relabel}
\end{figure}

QRC embeddings, being generated from quantum dynamics, are potentially hard to obtain with classical numerical simulation. Thus, the data transformation induced by QRC might have unique qualities. To quantify this, we consider the QRC action as a kernel that maps the feature vector space to embedding vector space, and we apply the recent idea of kernel geometry difference \cite{Huang2021}. Considering two kernels, in this case, quantum ($K_q$) and classical ($K_c$), if the geometric action of the kernels differs significantly, it is possible to transform the dataset labels such that the synthetic dataset thus obtained can realize the performance difference between the kernels. This synthetic data construction follows from the leading singular vector ($\mathbf{v}_{\infty}$) of the kernel geometry distance matrix, $g_{cq}$:    $g_{cq}^2=\sqrt{K_q}K_c^{-1}\sqrt{K_q}$. In practice, we assign the new label to the data according to the values of $\sqrt{K_q}\mathbf{v}_{\infty}$ \cite{Huang2021}. For a binary classification task, the result is a synthetic dataset with the same features as the original but with labels permuted. One can then re-train and test the QRC and CRC pipeline with the given synthetic dataset.

We apply this synthetic data construction to the 3/8 binary MNIST classification task. To mitigate overfitting to sampling noise, we use half of the experimental measurements to construct the synthetic data, while the remainder is used for training and inference. On this synthetic dataset, we observe comparative quantum kernel advantage with a test accuracy difference of $(7.1 \pm 3.1) \%$, as shown in Fig.~\ref{fig:Fig4_relabel}(a). We note that the kernel advantage observed here is comparative, realized in the direct comparison between the two kernel methods. One can similarly construct a synthetic dataset where the classical kernel can outperform the quantum kernel. Nonetheless, this shows the existence of a dataset, where if one is given the dataset and runs through the QRC and CRC pipeline, QRC will exhibit comparative quantum kernel advantage. We demonstrate successfully that this QRC advantage can be observed in experiment.

We also examine how the number of measurements impacts the comparative quantum kernel advantage in both numerical simulations and experiments, as shown in Fig.~\ref{fig:Fig4_relabel}(b). Similar to the trend observed in the original data, the experimental results closely mirror the simulation, albeit with a slight offset. Remarkably, the comparative quantum kernel advantage is evident even with $N_s=40$ in the experiment. This is significant compared to the typical $N_s \sim 1000$ required to achieve performance plateaus with original datasets, thus providing at least an order of magnitude reduction in runtime to achieve competitive QRC performance. Additionally, we construct another synthetic dataset using samples from noiseless QRC simulations. In tests with this dataset, the QRC experiments continue to outperform the CRC, exhibiting a robust kernel advantage. This further implies that experimental QRC kernels closely resemble their noiseless counterparts, underscoring the crucial role of quantum effects in achieving a comparative quantum kernel advantage.

\section*{Discussion and outlook}

Our work demonstrates a quantum reservoir computing algorithm on a neutral-atom analog quantum computer, and the algorithm is shown to be general-purpose, quantum-resource frugal, noise-robust, and scalable. We performed extensive experiments on hardware and numerical simulations to prove the algorithm concept, followed by scaling up successfully to systems of up to 108 qubits, substantially advancing the scale of previously reported QML experiments. The potential classical intractability of the QRC transformation is quantified through kernel geometry, and comparative quantum kernel advantage is observed in experiments.

The findings from our study open several avenues for exploration, improvement, and generalization. The most straightforward enhancements could involve scaling up both the experimental sampling rate and system size, as our results suggest substantial potential for performance gains. Besides further hardware advancements in neutral-atom analog quantum systems, improvements could also stem from tailoring the algorithm to different platforms, including distinct analog implementations~\cite{Shaw2024}, digital quantum computers, and even early fault-tolerant quantum platforms~\cite{ Bluvstein2023, Acharya2023, Mayer2024, Self2024}. The physical insights we provide on the universal parameter regime and noise robustness are likely to translate well across different quantum modalities. More fundamentally, QRC could be employed as a versatile QML tool: its combination of QRC kernels with classical postprocessing mirrors the function of the widely explored variational quantum circuits in machine learning applications, but with significantly lower quantum resource requirements. Extensions to the QRC architecture itself are another possible route, although potentially with more resource demands. For example, integrating recurrence into the architecture~\cite{Bravo2022} could enhance its utility for time series prediction, akin to traditional classical reservoir approaches~\cite{Jaeger2004}. The general-purpose nature of the algorithm allows strong hybridization with classical machine learning paradigms beyond supervised prediction, such as generative and unsupervised machine learning tasks. Moreover, our results demonstrate that certain datasets can exhibit comparative quantum kernel advantage, prompting further investigation into how to identify such datasets and determine which types of data are most conducive to leveraging quantum reservoir embeddings. This will be an important area of future research. A natural place to search for the advantage are explicitly quantum tasks, such as learning the phase diagrams of a Hamiltonian based on the quantum state or measurement outcomes~\cite{Huang2022}. Beyond that, an ostensibly classical dataset might have an implicit quantum nature, such as the activity of drug candidates~\cite{Vakili2024}.


\textit{Code  availability.---}A tutorial for reproducing proof-of-concept simulations and experiments is available at \url{https://github.com/QuEraComputing/QRC-tutorials}.

\textit{Acknowledgements.---}We acknowledge Boris Braverman and Jesse Amato-Grill for their early contributions to the build of Aquila. We acknowledge fruitful discussions with Casey Duckering, Tarushii Goel, and members from Mikhail Lukin's group. This work is supported by the DARPA IMPAQT program (grant number HR0011-23-3-0009), DARPA-STTR award (Award No.~140D0422C0035), and the DARPA ONISQ program (grant number W911NF2010021). H.-Y.H. and S.F.Y. acknowledge funding by the NSF through the Qu-IDEAS HDR program (OAC-2118310). R.A.B. acknowledges funding from NSF through the GRFP. X.G. acknowledges support from NSF PFC grant No.~PHYS 2317149, start-up grants from CU Boulder, U.S. Department of Energy, Office of Science, National Quantum Information Science Research Centers, and Quantum Systems Accelerator. The numerical studies were performed on the high-performance computing system Perlmutter, a NERSC resource, using NERSC award DDR-ERCAP0030190. This research was developed with funding from the Defense Advanced Research Projects Agency (DARPA). The views, opinions, and/or findings expressed are those of the author(s) and should not be interpreted as representing the official views or policies of the Department of Defense or the U.S. Government. 

%

\clearpage
\onecolumngrid
\begin{center}
{\bf Supplementary Information: ``Large-scale quantum reservoir learning with an analog quantum computer''}
\end{center}
In this Supplementary Information, we present the details of experiments and calculations presented in the main text \ref{sec:calcdetails}, additional numerical simulation and experimental results for MNIST classification \ref{sec:MNIST},  Santa Fe laser timeseries prediction \ref{sec:timeseries}, and diseased tomato leaves classification \ref{sec:Tomato}. We describe several aspects of practical QRC implementations, including the description of universal parameter regime \ref{sec:parameters}, experimental consistency \ref{sec:consistency}, and the encoding performance hierarchy discussion \ref{sec:hierarchy}.

\section{Experimental and calculational methods}
\label{sec:calcdetails}

\subsection{Data encoding}

\subsubsection{Feature extraction and dimension reduction}

Quantum reservoir computing (QRC) has recently emerged to be a promising approach suitable for near-term quantum hardware~\cite{Fujii2017, Ghosh2019, Ghosh2021, Martinez2021, Saeed2021, Suzuki2022, Burgess2022, Bravo2022, Dudas2023, Yasuda2023, Gotting2023, Xia2023, Mujal2023, Hu2023, Senanian2023, ahmed2024, Sannia2024}. The QRC algorithm we describe here is co-designed for neutral-atom quantum hardware. In our protocol, the encoding of the data into a QRC pipeline starts with a feature extraction and dimension reduction step that is dataset-dependent. In the case of MNIST data \cite{Deng2012}, the grayscale MNIST images are flattened to a vector whose components are normalized to $[0, 1]$ interval. The train data is used to fit the principle component analysis (PCA)  model \cite{Pearson1901, MultivariateStatsjl} up to a set model dimension, and the model is employed to extract PCA features for both the train and test data. The resulting PCA features are shifted and scaled such that they fit $[0, 1]$ interval.

In order to facilitate quantum hardware experiments, the binary MNIST classification tasks were performed with 1000 train and 200 test data each, while the 10-class experiment used 1500 train and 500 test data. The data features for the Santa Fe laser timeseries \cite{Weigend1993-bg} prediction were extracted from the raw timeseries by sliding windows of a set width. The outcomes for each window were the timeseries values at the first timepoint after the window. The experiments were performed with the training set derived from the time points between 1000 and 1400 and the test set from time points 1401 to 1485. Finally, the tomato dataset was prepared by choosing three classes of tomato leaf images from the plant village dataset \cite{hughes2016tomato}, for a total of 498 images, 400 of which represented the training data. The images were converted to grayscale and scaled to a uniform size of 256$\times$256 pixels. Depending on the dimensions of the features for a specific QRC experiment, the images were then downscaled to the appropriate resolution. The downscaling corresponds to taking the intensity average over a box of the size of the downscaled image pixel with the center at the downscaled pixel position, thus suppressing aliasing. The final downscaled image was prepared for encoding by normalizing the pixel intensities to $[0, 1]$ interval.

\subsubsection{Encoding to Rydberg Hamiltonian dynamics}

The encoding of the data into the Rydberg Hamiltonian parameters and dynamics, in general, follows the principles of the universal parameter regime described in detail in the Sec.~\ref{sec:parameters}. Thus, all energy scales including quantum mixing (Rabi frequency, $\Omega$), entangling (nearest neighbor interactions, $V_{\mathrm{n. n.}}$), encoding (for example, $\Delta_{\mathrm{l}}^{\mathrm{max}}$, local detuning amplitude), and probing ($1/\Delta t$, where $\Delta t$ is the evolution timestep) are designed to be of comparable scale.

The position encoding of the binary and 10-class MNIST experiments was performed with $\lambda = 3$ and $d_0=10 \, \mu \mathrm{m}$, where $\lambda$ is the encoding scale and $d_0$ is the nearest neighbor atom distance before encoding. In both experiments, 8 PCA components are encoded into 8 nearest neighbor interactions of a 9-qubit chain, according to $V_{i,i+1}=V^{(0)}(1+\lambda x_i)$, with $x_i$ being the feature vector. 
In everything that follows, the atom positions were rounded to two decimals of precision. The Rabi frequency during each quantum dynamics instance was ramped up from zero in $0.05 \, \mu \mathrm{s}$ to a constant maximum value, $\Omega= 2\pi \, \mathrm{MHz}$, and then ramped down to zero in $0.05 \, \mu \mathrm{s}$; global detuning was kept constant at $\Delta_{\mathrm{g}} = 2\pi \, \mathrm{MHz}$. The pulse shape thus respects the constraints of the quantum hardware \cite{Wurtz2023}.  Given that the parameters of the Rydberg Hamiltonian implemented in Aquila \cite{Wurtz2023} correspond to Rydberg interactions given by $V_{ij}=C/\Vert \mathbf{r}_i - \mathbf{r}_j\Vert^6$ with $C=862690 \times 2\pi \, \mathrm{MHz} \, \mu \mathrm{m}^6$, the initial effective dimensionless interaction strength parameter (initial ``blockade radius'' \cite{Bernien2017}) $R_{b}^{(0)}/a=[C/(\Omega d_0^6)]^{(1/6)} \approx 0.98$. The Hamiltonian dynamics was probed at 5 successive timesteps, each lasting $\Delta t= 0.5 \, \mu \mathrm{s}$. For each timestep, a number of experimental repetitions were performed to generate measurement shots for embedding calculation. This number of repetitions and, thus, hardware runtime, was reduced by performing quantum dynamics in parallel on 6 identical parallel chains, with an interchain distance of $15 \, \mu \mathrm{m}$, allowing for practically isolated chain dynamics due to the fast decay of Rydberg interaction. All of the numerical simulations from the main text were performed for the same parameters as reported here with the assumption of ideally isolated chains using the Bloqade package \cite{bloqade2023quera}.

The Santa Fe laser dataset was probed with all three encodings. The position encoding was applied for 11 datapoint wide window features and correspondingly 12-qubit chains, with all the parameters equivalent to the ones already described with 10-class MNIST experiments, besides smaller $\lambda=1.5$ due to the increased mean dynamical range of the Santa Fe laser data. The local detuning encoding was done with 12-qubit chains and 12-wide window features with a constant nearest neighbor distance of $d=10 \, \mu \mathrm{m}$ and the same Rabi pulse and probing scheduled as described for the MNIST experiments. The global detuning was kept constant at $\Delta_\mathrm{g}= 4 \, \mathrm{MHz}$, while the local detuning was constant in time with the site-dependent amplitudes encoding the data features as $\Delta_\mathrm{l}[i]=\Delta_{\mathrm{l}}^{\mathrm{max}}x_i$, with $\Delta_{\mathrm{l}}^{\mathrm{max}}=-8 \, \mathrm{MHz}$. Finally, the global pulse encoding was also performed on the 12-qubit chains with 10-wide window features. The chain geometry was fixed but irregular according to the nearest neighbor distances being generated as $(8.9-g_i)\, \mu \mathrm{m}$, with $g_i$ being a vector with components drawn as uniform random numbers on $[0, 1]$ interval. The irregular chain geometry was found to benefit the global pulse encoding performance, as described in the Sec.~\ref{sec:timeseries}. The Rabi pulse was the usual $0.05 \, \mu \mathrm{s}$ ramp-up and ramp-down to a maximum of $\Omega= 3\pi \, \mathrm{MHz}$ (see Fig.~\ref{fig:Fig2_timeseries}). The probing schedule consisted of 10 probing timesteps each of $\Delta t = 0.35 \, \mu \mathrm{s}$, which coincided with the global pulse encoding schedule. The encoding consisted of 10 window features being rescaled to the global detuning range of $[0, 12]\, \mathrm{MHz}$ and then connected with 9 piecewise linear segments. The last global detuning segment was constant, corresponding to the value of the final data point in the window. An example of pulse protocols is shown in Fig.~\ref{fig:Fig2_timeseries} of the main text. All three encodings employed the 6-parallel chain construction, and simulations quoted in the main text corresponded to the same parameters.

The tomato leaf images dataset was encoded with 2D position encoding. Due to the minimum row distance constraint of the hardware \cite{Wurtz2023}, the irregular row positions that would result from position encoding in the vertical direction were avoided, and thus only horizontal nearest neighbor interactions encoded the pixel data. The result is that $(R_x-1)\times R_y$ pixel image was encoded into the $R_x \times R_y$ qubit array with position encoding $\lambda =4.0$ and $d_{0x}=d_{0y}=10 \, \mu \mathrm{m}$. The maximum Rabi amplitude during a constant drive with $0.05 \, \mu \mathrm{s}$ ramps was $\Omega=5\pi \, \mathrm{MHz}$, the same as the constant global detuning $\Delta_{\mathrm{g}}$. The increased $\Omega$ motivated the corresponding increase in the encoding and entanglement scales with $\lambda=4.0$, as well as the quantum dynamics probing interval of $\Delta t=0.3 \, \mu \mathrm{s}$ with 5 sampling steps. Thus, an overall faster protocol was performed to facilitate coherence in a large system of qubits. While the 2D arrays of $7 \times 7$ atoms and larger could not be run with more than one array in parallel due to the limited accessible physical space on the hardware, the sampling in the smaller arrays was facilitated by parallelization with $15 \, \mu \mathrm{m}$ minimum distance between any two atoms of different parallel arrays. The simulations were performed with the same parameters as experiments for the accessible small systems.

\subsection{Data postprocessing and training}

\subsubsection{Embedding calculation}

The quantum reservoir dynamics probed on hardware resulted in the set of bare experimental measurement shots for each data and the probing time. Not all of the experimental sequences started with the perfectly sorted array of qubits \cite{Wurtz2023}. In the 1D chain experiments, the shots from imperfect initial sorts were discarded from the calculation as the sorting imperfection affects the qubit connectivity. In all the 1D system sizes explored, this resulted in less than 10\% of discarded shots and thus had a negligible performance impact. In the case of the 2D QRC, the point defects introduced by imperfect sorting do not strongly affect the connectivity of the system, and thus no sample postselection was made in the results reported. At the largest system sizes, around 60\% of the shots would have a sorting defect, and thus, the significant sampling reduction would negatively impact the performance. No other postprocessing was applied to experimental shots. The local observable expectation values were then estimated. In all cases, they included all of the $\langle Z_i \rangle$ expectation values. In the proof-of-principle 1D chain experiments, all of the $\langle Z_i Z_j \rangle$ with $i<j$ were also used, while the large-scale 2D experiments employed only the nearest-neighbor connected subset of the two-point correlators ($\langle Z_i Z_j \rangle$ where $j$ is a nearest neighbor of $i$ and $i<j$). The flattened vector of local observable expectation values forms the QRC embeddings, $u_i$. The reduced set of correlators of the large system limited the embedding dimension to a scale similar to the dataset size, thus improving the error estimates (see Sec.~\ref{sec:MNIST}) without significant performance impact in practice. The same embedding calculation procedure was used in the finitely sampled simulation, while the exact simulation calculated the embeddings directly.

\subsubsection{Classical models}

With embeddings generated, the next step in the QRC algorithm involved training the classical model on the training set and evaluating performance on the test set. In all of the classification tasks, the classical model employed was linear support vector machine (SVM), with the C-SVC algorithm \cite{Libsvm2011}. The optimal soft-margin regularization parameter, $C$, was found via grid search using an 80/20 split of the training set into training and validation in the one instance of dataset permutations (see later discussion) and was kept the same in all the other instances. In practice, the optimal results were very weakly dependent on $C$ close to $C=1$. The timeseries prediction was trained using the linear support vector regression (SVR), with the $\epsilon$-SVR algorithm \cite{Libsvm2011}. The hyperparameters $C$ and $\epsilon$ were again optimized by the grid search with 80/20 training/validation split of the train set for one instance of the training size (see later discussion) and were kept the same in the other instances.

The performance of the QRC was compared with several classical models. The simplest of these represented the QRC pipeline without the quantum dynamics and embeddings, using the features that are otherwise encoded in the Hamiltonian. The comparison was done with the linear SVM/SVR, and they included the same training and hyperparameter search procedures as for the QRC itself. 

As an example of a classical non-linear method, we employed a 4-layer feedforward neural network on the features encoded otherwise in the Hamiltonian. The neural network had an input layer the same as the feature dimension, two hidden layers, each with 100 nodes followed by a ReLU activation function, and an output layer that was a softmax layer for classification and a single neuron for the regression tasks. The network was trained via gradient descent using the Flux library \cite{Flux2018}. The training hyperparameters, including L1 regularization, learning speed, and the number of epochs, were optimized in the same procedure as the SVM/SVR hyperparameters described earlier. Another classical non-linear method was employed for some of the kernel geometry construction calculations, i.e., the Gaussian (radial basis function, RBF) kernel. The implementation employed the C-SVC algorithm \cite{Libsvm2011} and included the optimization of $C$ and $\gamma$ hyperparameters with the same above-described procedure. 

Finally, we also compared the QRC with the analogous classical reservoir (CRC). The CRC is derived by promoting all the qubits (spin $1/2$'s) to $S\rightarrow\infty$, thus making them classical unit vectors ($\hat{S}$). The dequantized dynamics of the Rydberg Hamiltonian is described by:
\begin{equation}\label{eq:CRCdynamics}
    \frac{\mathrm{d} \hat{S}_i}{\mathrm{d} t}= \frac{\partial H[\hat{S}]}{\partial \hat{S}_i} \times \hat{S},
\end{equation}
where the effective instantaneous magnetic field acting on the $\hat{S}_i$ is:
\begin{align}\label{eq:CRCfield}
    \frac{\partial H[\hat{S}]}{\partial \hat{S}_i}=&\frac{\Omega(t)}{2} \hat{x}\cr&+ \left[-\frac{\Delta_i(t)}{2} + \frac{1}{4} \sum_{j\neq i}V_{ij}\left(1 + \hat{S}_j^{(z)} \right)\right] \hat{z}.
\end{align}
Thus, the dynamics of $N_q$-classical spins can be efficiently simulated by integrating a system of $3N_q$ equations. The QRC-equivalent CRC is performed with the same time profiles of pulses and the values of all the parameters, as described for the QRC pipeline in previous sections. The CRC embeddings generated are derived from the $z$-axis projections of the spins at the probing time and their products. Depending on the task, the training and inference were done through the same linear SVM/SVR models as described for the QRC. The space of all the classical spin configurations has a one-to-one correspondence to all quantum product states, and we can thus consider CRC as the entanglement-less limit of the QRC.

\subsubsection{Uncertainty estimation}

The stochastic nature of the QRC and of the training introduces uncertainty into each estimation of the model performance. In order to provide an unbiased estimate of the mean performance uncertainty, we performed averaging over multiple instances of training and inference, according to the following:
\begin{itemize}
    \item In order to capture the uncertainty in quantum embeddings due to finite sampling, we included 5 instances of resampled embeddings for a dataset. The resampling was performed by randomly choosing 90\% of the shots from the whole shot pool to recalculate new embeddings for each training instance. This step was inapplicable in the case of exact simulation or for any of the classical comparison models.
    \item The variability induced by the finite dataset size was taken into account by either dataset permutation or training set resampling, depending on the tasks. For the classification tasks, the 5 separate instances were drawn by permuting the dataset order and making a new train/test split for each of the permutations. For the timeseries prediction tasks that have causal structure, 5 different instances of the training set were made by randomly discarding 10\% of the training data windows in each. 
\end{itemize}
The full set of instances for each of the test accuracy datapoints presented thus consisted of 25 instances, each with a different resampling of the experimental shots and permutation/resampling of the dataset. The training and inference were then repeated according to the optimal model hyperparameters found on the validation instance, as described previously. The final reported test accuracy is the mean of the test accuracy on all of the instances, while the reported error is the standard deviation of the same test accuracy instance set. The equivalent instance resampling procedure was performed for all the classical models probed, with the exception of the embedding resampling that did not apply in this case.  We have found that the reported results are stable against further increases in the number of resampling instances and robust against both sizeable hyperparameter changes and the change of the dataset resampling strategy from permutation to training set resampling. In the cases where data apparently lacked error bars, the error estimates are below the sizes of the plot markers, which is typical for the exact simulations of large datasets (see Supplementary Information). On top of the above mentioned instances, additional training and inference instances were generated for kernel geometry calculations due to the regularization procedure employed, as described in the following section. 

\subsection{Kernel geometry construction}

Kernel geometry construction calculations were performed according to the prescription in Ref.~\cite{Huang2021}, with new protocols devised for the specifics of experimental data. The calculations started by normalizing the embedding vectors, $\hat{u}[n]=\mathbf{u}[n]/\Vert \mathbf{u}[n] \Vert$. In the case of the QRC and CRC kernels (see Fig.~\ref{fig:Fig4_relabel}), the normalized embeddings are then used to directly calculate the kernel matrices, with $K_{nm}=\langle \hat{u}[n], \hat{u}[m] \rangle$, where $n (m)$ enumerate the data. For comparison with another common classical non-linear kernel, we used the Gaussian (radial basis function, RBF) kernel (see Supplementary Information), derived from the normalized feature vectors $\hat{x}[n]$ as $K_{nm}=\exp{-\gamma\Vert \hat{x}[n] - \hat{x}[m] \Vert^2}$, where $\gamma$ is a hyperparameter to be optimized. As a result, for each quantum-classical kernel pair, we obtain $K_c$ and $K_q$ kernel matrices and can thus proceed to analyze the spectrum of the $g_{cq}^2=\sqrt{K_q}K_c^{-1}\sqrt{K_q}$ matrix. While kernels are positive-semidefinite and the square roots are well-defined, the inverse of the classical kernel can be ill-defined. To ameliorate this, we turn to the regularization procedure described in Ref.~\cite{Huang2021}, considering instead the regularized geometry matrix instead:
\begin{equation}\label{eq:reggcq}
    g_{cq}^2[\delta]=\sqrt{K_q}\sqrt{K_c}(K_c+\delta I)^{-2}\sqrt{K_c}\sqrt{K_q},
\end{equation}
where $\delta$ is the regularization parameter. The lowest relevant $\delta$ is the lower end of the non-zero singular values of $K_c$, while the upper reasonable bound should still be $\delta \ll 1$, such that the norm of $K_c+\delta I$ does not appreciably differ from $K_c$. We find that in all of the $K_c$ cases we considered, this is practically realized with $\delta \in [10^{-8}, 10^{-2}]$. 

The results should ideally be weakly dependent on the regularization. In order to assure this, we performed the kernel-based synthetic data construction procedure independently for each $\delta$, sampled from a geometric series with $10^{1/4}$ ratio in the $[10^{-8}, 10^{-2}]$ interval. The singular vector of the leading singular value of $g_{cq}^2[\delta]$, $\mathbf{v}_{\infty}[\delta]$, is used to construct labels for a synthetic balanced binary classification task ($\mathbf{y}'$), according to \cite{Hu2023}:
\begin{equation}\label{eq:newlabels}    \mathbf{y}'[\delta]=\mathrm{sign}\left[\sqrt{K_q}\mathbf{v}_{\infty}[\delta]-\mathrm{median}\left(\sqrt{K_q}\mathbf{v}_{\infty}[\delta]\right)\right].
\end{equation}
While $\sqrt{K_q}\mathbf{v}_{\infty}$ might be used already to construct the most quantum-native synthetic regression task, we favor the binary classification task due to its performance interpretability. In the case where the original task was a balanced binary classification, the procedure thus simply permutes the labels of the classes. 

Once the synthetic datasets are constructed, the quantum and classical kernel matrices are used once again for training and inference by employing the kernel-specific C-SVC algorithm \cite{Libsvm2011} in the same fashion as described previously. The only appreciable difference is that the train/test splits in the case of the binary classification were changed to 800/400 from the usual 1000/200, in order to lower some of the dataset size-related uncertainty. In the case when the QRC embeddings were drawn from finitely sampled classical simulation or quantum experiment, the original kernel geometry construction requires modification to avoid overfitting synthetic data to a particular realization of the sampling noise. This is achieved by taking half of the experimental shots at each data point in a set to form the kernel $K_{q1}$ used for the calculation of new labels. The second, non-overlapping part of the shots is used to calculate kernel $K_{q2}$ and perform training and inference. Alternatively, $K_{q1}$ can be constructed from finitely sampled or exact simulation at scales where the simulation is available. The experimental and simulation-based synthetic dataset construction were applied in Fig.~\ref{fig:Fig4_relabel}(b) of the main text, as presented with two orange data series. The comparative quantum kernel advantage was observed in experimental test performance for both synthetic datasets.

The result of the kernel geometry construction calculations at each regularization value $\delta$ are quantum and classical test accuracies. We confirm that, as expected, the qualitative relation between the two is independent of the regularization parameter, as shown on an example in Supplementary Information for the exact simulation of the 10-class MNIST dataset with QRC-CRC comparison. The test accuracy difference is also only weakly dependent on the regularization, even though regularization affects individual method accuracies. To incorporate the regularization-induced uncertainty, we include synthetic data results at different $\delta$ into the set of instances used for estimating test accuracy and combine it with the already described shot and dataset resampling instances. Thus, the effect of regularization is mostly seen as increased uncertainty in the test accuracy difference results.

Finally, while CRC and QRC kernels are parameter-free, the Gaussian kernel includes a tunable hyperparameter $\gamma$. The hyperparameter was optimized separately at each $\delta$ via grid search, such that the whole kernel geometry calculation (both synthetic data construction and training/inference) results in the smallest possible QRC-Gaussian test accuracy difference, thus providing the best chances for the Gaussian kernel in the synthetic scenario (see Supplementary Information), hence serving as a robust test for the comparative quantum kernel advantage. It is significant to note that the kernel procedure is evidently not symmetric with respect to classical/quantum kernel order. The comparative quantum kernel advantage observed for synthetic data derived from $g_{cq}$ matrix that favors the quantum method does not have to hold for the $g_{qc}$ derived synthetic data. This does not preclude a distinct possibility of quantum-suitable datasets hinted at by the comparative quantum kernel advantage we observe.

\section{Supplementary results for handwritten digits classification}
\label{sec:MNIST}

Here we provide details of the QRC performance on the MNIST classification dataset. The large part of the section deals with extensive numerical simulations of the QRC on the 10-class MNIST data with a larger piece of dataset processed than in the experiments. These simulations show the effect of system size scaling and sampling in Sec.~\ref{sec:systemsize}, provide a description and the numerical example of the universal parameter regime in Sec.~\ref{sec:parameters}, and provide additional evidence of comparative quantum kernel advantage in Sec.~\ref{sec:relabeledperf}. Later, we provide additional details of the experimental performance, including the discussion of experimental consistency tracking in Sec.~\ref{sec:consistency} and the 0/1 binary MNIST classification results in Sec.~\ref{sec:01binary}.

\subsection{System size scaling and sampling}
\label{sec:systemsize}

\begin{figure*}[htb]
\centering
\includegraphics[width=0.75\textwidth]{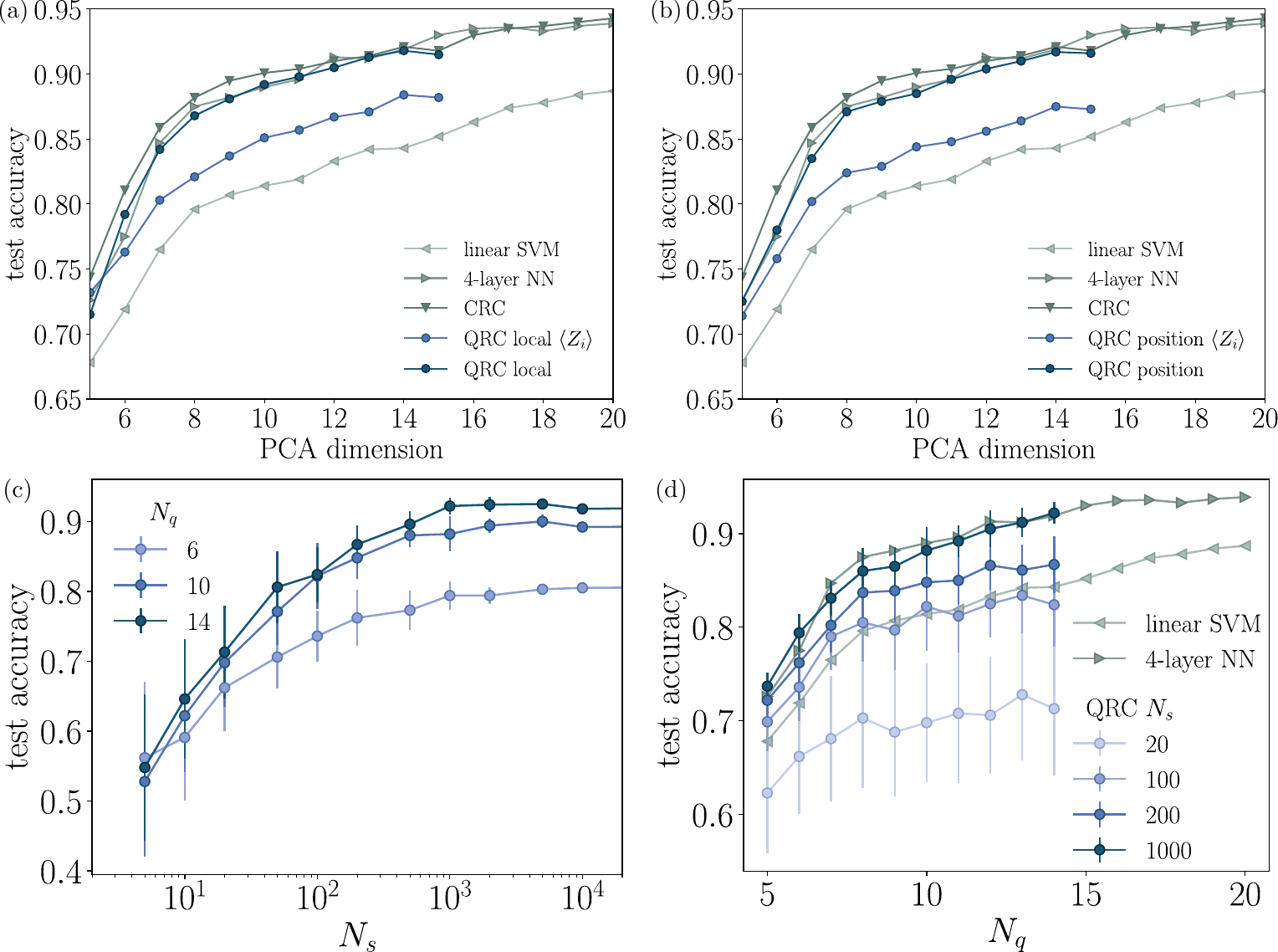}
\caption{\textbf{Proof of concept neutral-atom QRC implementation in numerical simulations.} The 10-class MNIST handwritten digit classification dataset was used to gauge QRC performance in numerical simulations. The image data was downsampled with principal component analysis (PCA) and encoding through local pulse and position encodings to a linear neutral-atom chain. (a) Test classification accuracy of the local pulse encoded QRC and several classical models as a function of PCA encoding dimension (or equivalently, qubit number, $N_q$). Comparison between QRC that uses both single- and two-site- local operator expectation values and single-site-only approach is also presented. 
(b) The test accuracy of the position-encoded QRC and several classical methods as the function of the PCA encoding dimension. (c) The scaling of local pulse encoded QRC with the number of shots drawn per datapoint ($N_s$) for several system sizes. (d) Test accuracy of local pulse encoded QRC as a function of qubit number for several $N_s$.
}
\label{fig:FigS2_MNISTscaling}
\end{figure*}

In order to evaluate the potential of the QRC method beyond the scope of experiments, we perform extensive numerical simulations on the MNIST dataset. The size of the dataset processed with these simulations was 10000 train and 1000 test data samples, with the same QRC and classical model pipeline as described in Sec.~\ref{sec:calcdetails}, with the exception of the encoding scale factor of $\lambda=1.0$ for the position encoding. The main results are shown in Fig.~\ref{fig:FigS2_MNISTscaling}. Throughout the PCA dimensions simulated, the QRC consistently matches the test accuracy curves of the nonlinear classical methods and significantly outperforms the linear SVM. It is likely that all nonlinear methods, in this case, reach the performance thresholds accessible given the data size and the number of PCA components used. Furthermore, the QRC that only uses a one-site observable set, $\langle Z_i \rangle$ for embeddings, is not able to achieve the same performance, being between the linear SVM and the nonlinear thresholds. We have also checked that adding three-point correlators does not lead to detectable performance benefits, both in the simulations and experiments. Comparing the local and position encodings in Fig.~\ref{fig:FigS2_MNISTscaling}(a) and (b) respectively, we observe equivalent performance both with one- and two-site correlator embeddings, which we ascribe to the partial mapping between the two encodings, as described in the Sec.~\ref{sec:parameters}. 

The most significant practical limitation to the QRC performance is the number of experimental shots drawn for each embedding. The performance scaling with the number of shots drawn in MNIST dataset simulations is shown in Fig.~\ref{fig:FigS2_MNISTscaling}(c) and (d). The sampling overhead observed stems from the $1/\sqrt{N_s}$ uncertainty due to the shot noise in the embeddings and is thus only polynomial due to our choice of local observable embeddings. Furthermore, we typically observe that the performance plateau at which the finite number of shots matches the exact simulation. For the 10-class MNIST dataset, this is typically at $N_s \sim 10^3$ and does not depend on the system size; the plateau point is mostly likely determined by the classification thresholds and the intrinsic noise of the dataset. In general, practical datasets have a level of intrinsic noise due to data or labeling imperfections, and a finite sample plateau is expected to be observed, albeit at different $N_s$. More forgiving classification thresholds in the simpler classification tasks, such as binary classification, are expected to additionally shift the plateau position to lower $N_s$. Beyond the plateau, the system size scaling curves are similar at different sampling scales [see Fig.~\ref{fig:FigS2_MNISTscaling}(d)], further supporting the scalability of the algorithm.   

\subsection{Universal parameter regime}
\label{sec:parameters}

Constructing a truly gradient-free QRC algorithm requires a choice of Hamiltonian dynamics parameters that allow good performance that is largely independent of the dataset. In the extensive numerical simulations on the datasets probed, we consistently observe the same parameter regime leading to a near-optimal performance with a broad optimum point. This broad universal regime can be understood physically by considering the following energy scales:
\begin{itemize}
    \item mixing scale, determined by the effective maximum Rabi amplitude, $\overline{\Omega}$;
    \item entanglement scale, determined by the average Rydberg interaction strength in the array, $\overline{V}$;
    \item encoding scale, which, depending on the encoding method, can include average local detuning (local pulse encoding), fluctuating part of the interaction strength (position encoding), or average global detuning encoding carrying amplitude (pulse encoding), $\overline{\lambda}$;
    \item probing frequency scale, the inverse of the quantum dynamics probing time, $(\overline{\Delta t})^{-1}$;
    \item encoding frequency scale, specific to global pulse encoding, being the inverse of the datapoint encoding interval, $(\overline{\Delta \tau})^{-1}$.
\end{itemize}
The global parameter regime can be thus described as the rough equivalence of all relevant energy scales:
\begin{equation}\label{eq:unparam}
    \overline{\Omega} \sim \overline{V} \sim \overline{\lambda} \sim \left(\overline{\Delta t}\right)^{-1} \sim \left(\overline{\Delta \tau}\right)^{-1},
\end{equation}
where in practice, the broadness of the optimal regime allows up to between a quarter and half a decade of difference between the ratios of different scales.

Similar to related recent works~\cite{Martinez2021, Lu2024}, we can understand this regime of comparable scales via physical arguments. For example, if the mixing scale is significantly lower than the entanglement scale, the quantum dynamics will be constrained and very slow. In the opposite case, the interaction between the qubits that induce transformations, including several data features, will be insufficient; in both cases, the resulting QRC embeddings will not be as effective. Similarly, the encoding scale that is significantly lower than other scales will fail to express the data, while the dominant encoding scale approaches the classical limit in which no QRC transformation is performed. Probing and encoding frequencies that are significantly higher than the Hamiltonian energy scales provide redundant embeddings that do not increase performance and can affect trainability due to the high embedding dimension or, in the case of a finite number of shots, provide less variety in embedding transformations. Very low probing and encoding frequencies, however, can lead to probing the data-independent observables due to thermalization (additionally decohered in the experiment), providing less effective embeddings than in the non-thermal regime. The universal parameter regime thus described is applied for the construction of all simulated and experimental QRC protocols, with good effect. Particularly significant is the success of the protocols constructed for the large 2D QRC tasks where no equivalent simulation could be performed. 

\begin{figure*}[htb]
\centering
\includegraphics[width=0.8\textwidth]{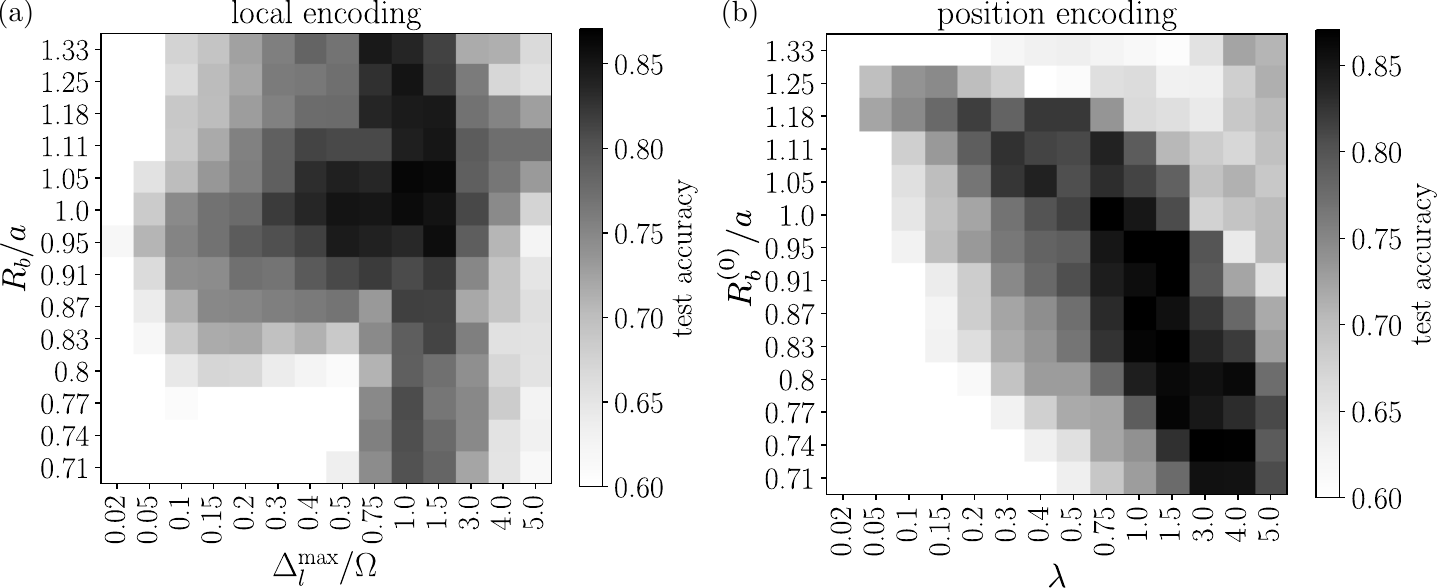}
\caption{\textbf{An example of universal parameter regime in numerical simulations of QRC.} (a) Test classification accuracy of local pulse encoded QRC on an 8-PCA 10-class MNIST task as a function of local pulse encoding scale ($\Delta_l^{\mathrm{max}}/\Omega$), and effective blockade radius ($R_b/a$). (b) Test accuracy of the position-encoded QRC on the same task as a function of position encoding scale ($\lambda$) and effective initial blockade radius ($R_{b}^{(0)}/a$).
}
\label{fig:FigS3_universal}
\end{figure*}

In Fig.~\ref{fig:FigS3_universal}, we provide examples of the numerically simulated parameter searches that showcase directly the universal parameter regime features. The MNIST test accuracy dependence on the encoding-mixing ratio and the dimensionless interaction-mixing ratio parameter $R_{b}/a=(V_{\mathrm{n.n.}}/\Omega)^{1/6}$ is presented in Fig.~\ref{fig:FigS3_universal}(a) for the case of local pulse encoding. The characteristic cross feature of the test accuracy is the consequence of the universal optimal parameter regime, as it is concentrated around both of the parameter ratios $\sim 1$. The size of the optimal region is about half a decade wide in both parameter ratios (note that the effective interaction-mixing ratio has a sixth root due to usual definitions). Related behavior can be observed in the features of Fig.~\ref{fig:FigS3_universal}(b), where the MNIST test accuracy is shown as a function of the bare ($R_{b}^{(0)}/a$) interaction-mixing ratio and the interaction-encoding ratio ($\lambda$) for the case of position encoding. Note that the increased interaction-encoding ratio $\lambda$ increases the actual interaction-mixing ratio, $R_{b}/a$, due to $V_{\mathrm{n.n.}}=V^{(0)}_{\mathrm{n.n.}}(1+\lambda x_i)$. As a consequence, the region of optimality spreads diagonally with increasing $\lambda$ requiring decreasing $R_{b}^{(0)}/a$ ratios to keep $R_{b}/a \sim 1$. The diagonal feature does not extend to low $\lambda$ and high $\lambda$, with around half of a decade of optimal performance possible around $\lambda \sim 1$, and similarly for the vertical width of the feature in terms of $(R_{b}^{(0)}/a)^6$. In practice, the position of the diagonal feature can be directly calculated from the average size of the data features, $x_i$, by requiring that the emergent average $R_{b}/a \sim 1$, thus circumventing simulations. 

Within their regions of optimality, local and position encoding are generally observed to have similar performance, as also noted in Sec.~\ref{sec:systemsize}. This can be understood from the position encoding partially inducing effective local detuning encoding, as becomes evident by rewriting the Rydberg interaction through $Z$ operators, using $Z_i=2n_i-I_i$, as follows:
\begin{equation}\label{eq:encconnection}
    \sum_{i<j }V_{ij} n_i n_j + \sum_{i}\Delta_i n_i \rightarrow \frac{1}{4}\sum_{i< j }V_{ij} Z_i Z_j - \frac{1}{2}\sum_{i}\left(\Delta_i+\frac{1}{2}\sum_{j, j\neq i}V_{ij}  \right)Z_i + \mathrm{const}.
\end{equation}
Thus, the position encoding modulation effectively induces local detuning modulation, with the local detuning modulation on one site being dominated by the features encoded at its nearest neighbors. The optimal parameter regimes follow a similar form in the $Z$-operator representation, as confirmed numerically. No similar connection can be found for the pulse encoding, and as a consequence, we observe a significant difference in performance in the optimal regime, as discussed in the main text and Sec.~\ref{sec:hierarchy}.

\subsection{Kernel geometry synthetic data performance}
\label{sec:relabeledperf}

 Here, we provide additional results establishing comparative quantum kernel advantage (CQKA) in MNIST dataset simulations, supplementing the reported CQKA results from the main text. Our results in simulations show that the CQKA is sizeable and robust with respect to different data (3/8 and 10-class MNIST), different classical kernels (CRC and Gaussian), regularization, and the system size.   
 
\begin{figure*}[htb]
\centering
\includegraphics[width=1.0\textwidth]{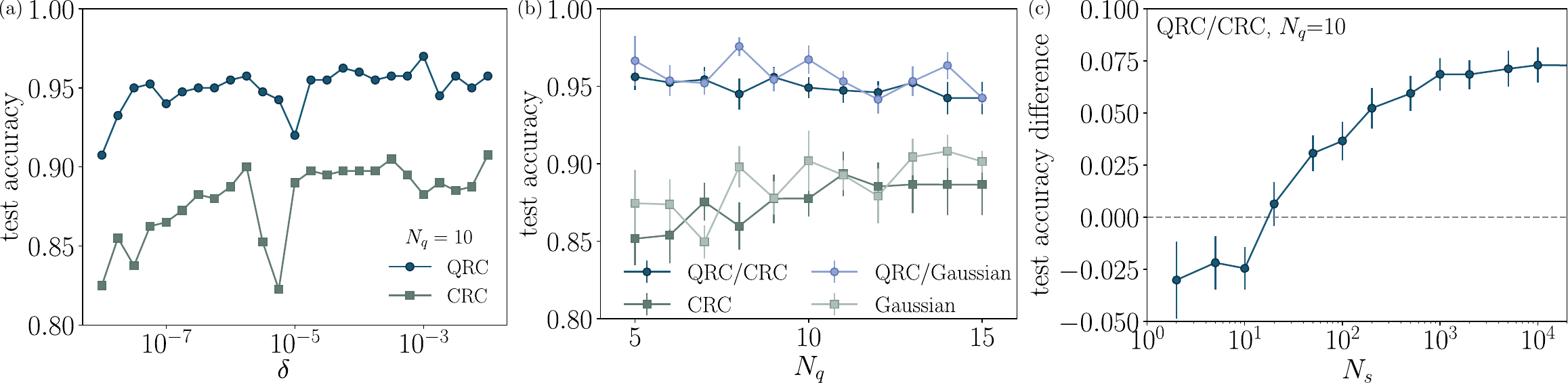}
\caption{\textbf{Kernel geometry construction results in numerical simulations of QRC.} (a) Test accuracy of QRC and CRC as a function of the regularization scale, $\delta$, in the synthetic MNIST task. (b) Test accuracies of the QRC kernel compared to two classical kernels, CRC and Gaussian kernels, as a function of qubit number, $N_q$ (or equivalent PCA dimension), in the synthetic MNIST task. Kernel geometry construction is done separately for two respective kernel pairs, QRC/CRC and QRC/Gaussian. (c) Test accuracy difference between QRC and CRC kernels as a function of the number of shots drawn per datapoint, in the 10-qubit synthetic MNIST task.  
}
\label{fig:FigS3_MNISTrelabeling}
\end{figure*}

 We start by establishing the weak regularization parameter dependence of the results on a typical example, shown in Fig.~\ref{fig:FigS3_MNISTrelabeling}(a), and covered in Sec.~\ref{sec:calcdetails}. Next, we expand the kernel geometry calculations to include both the QRC-CRC kernel comparison covered in the main text and the QRC-Gaussian kernel (see Sec.~\ref{sec:calcdetails}) comparison on a 10-class MNIST data. Fig.~\ref{fig:FigS3_MNISTrelabeling}(b) presents the test accuracy of the QRC kernel and the two respective classical kernels used for two separate kernel construction calculations -- CRC and Gaussian. Two comparisons result in two distinct synthetic datasets, and thus, the QRC test accuracy is reported separately for both as a function of the qubit number. In both quantum-classical kernel comparisons, sizable CQKA is observed, which is weakly dependent on the system size and saturates around $N_q=10$. The Gaussian kernel hyperparameter $\gamma$ was explicitly optimized to minimize this performance distance, but the CQKA magnitudes observed remained similar to the CRC comparison levels.

The test accuracy difference for one QRC-CRC example as a function of the number of shots drawn per embedding is shown in Fig.~\ref{fig:FigS3_MNISTrelabeling}(c). The shot scaling is similar to that observed for the original MNIST data, as described in Sec.~\ref{sec:systemsize}. While the performance plateau is still around $N_s \sim 10^3$, breakeven quantum-classical performance can be achieved with as low as $N_s=20$ shots. Thus, quantum-adapted data derived from kernel construction not only allows the detection of CQKA but also significantly lowers the sampling requirements for good QRC performance.

\subsection{Experimental consistency}
\label{sec:consistency}

\begin{figure*}[htb]
\centering 
\includegraphics[width=0.35\textwidth]{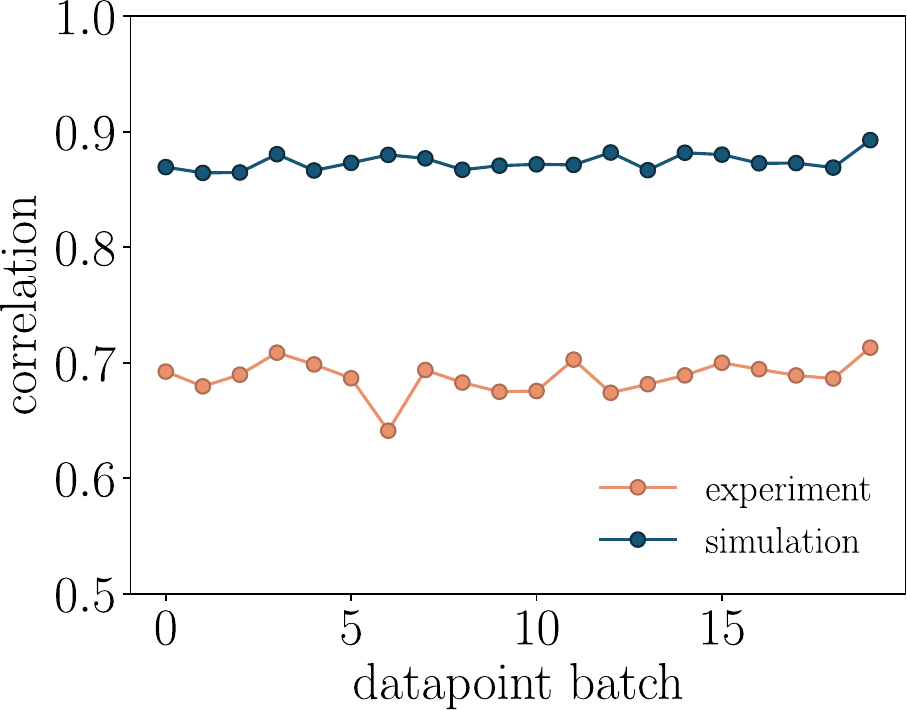}
\caption{\textbf{Experimental consistency during a QRC task.} The experimental consistency is tracked by statistical correlations between simulated and experimental embedding vectors on batches of data. Here presented are correlations between embedding vectors of the exactly simulated 3/8-binary MNIST task (see main text) and the embeddings drawn from the experiment (orange) and finite sampled simulation (blue). The number of shots drawn per datapoint is 110, while the total 1200 data used is divided into 20 batches of 60 that are used to calculate the embedding correlations.
}
\label{fig:FigS3N_consistency}
\end{figure*}

The most damaging source of experimental noise for the QRC performance is the systematic drift that results in distinct hardware runs of the same data providing inequivalent embeddings. In order to track and correct the hardware performance during the experimental runs, and quantify how close are hardware generated embeddings to the ones expected from simulation, we devise a simple measure: the statistical correlation between exactly simulated ($\tilde{\mathrm{u}}_i$), and experimental embedding vectors ($\mathrm{u}_i$), $\rho=\mathrm{corr}\left(\tilde{\mathrm{u}}_i, \mathrm{u}_i\right)$. In practice, the dataset is divided in batches and the embeddings from the whole batch are used in one flattened vector for the correlation evaluation. As an example of its use, the statistical correlation for finitely simulated and experimental embeddings as a function of the dataset batch is shown in Fig.~\ref{fig:FigS3N_consistency} for the 3/8 binary MNIST task.

From the presented results, it is clear that the experimental embeddings are very strongly correlated to the exact simulation, with a part of the correlation drop that can be explained with finite sampling and the rest due to noise effects. More critically for the QRC performance, the correlation is very weakly batch-dependent, at the same level as the finitely sampled simulation. The batch independence is expected generically for large enough batches, as each batch is a good representation of the dataset on average.  This consistency correlates strongly with the observed QRC performance. Any outliers observed in their correlation data were found to correspond to otherwise detectable hardware miscalibration events. In practice, such outlier batches were targeted for hardware reruns, resulting in improvements in the overall performance. The statistical correlation between the embeddings was thus a simple tool for ensuring practical hardware performance in QRC tasks. While the measure described here depends on the ability to provide simulated embeddings, a similar measure based on the correlation between different experimental batches could be employed for hardware consistency tracking without any additional hardware runtime overheads.

\subsection{MNIST 0/1-binary classification experiments}
\label{sec:01binary}

\begin{figure*}[htb]
\centering
\includegraphics[width=0.75\textwidth]{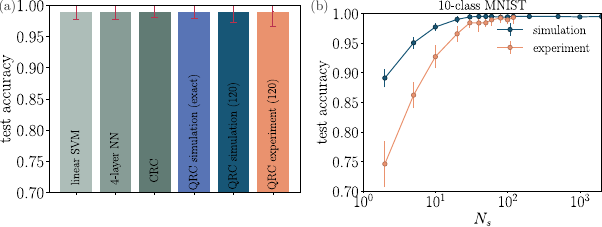}
\caption{\textbf{QRC performance on 0/1-binary MNIST classification task.} (a) The test classification accuracy of several classical machine learning methods and QRC on the 0/1-MNIST binary classification tasks. (b) The test accuracy of the experimental and simulated QRC as a function of number of shots drawn per datapoint, $N_s$, for the same task.
}
\label{fig:FigS3NN_binary01}
\end{figure*}

In addition to the experimental MNIST results provided in the main text, we have also performed the QRC experiments with the 0/1 binary classification task, with the same parameters as reported for 3/8 binary classification. The results are shown in Fig.~\ref{fig:FigS3NN_binary01}. The experimental QRC reaches test accuracy above 99\%. However, the same performance level is observed with all classical methods, including linear SVM baseline, pointing to the exceptionally wide classification margins for the task. This is further confirmed in the shot ($N_s$) scaling of the test accuracy, where the performance thresholds for the dataset are reached already with $N_s \approx 20$-$50$ in simulations and experiments.

\section{Supplementary results for timeseries prediction and encoding comparison}
\label{sec:timeseries}

Here we provide the details of the QRC simulations on the full Santa Fe laser dataset in Sec.~\ref{sec:fullSFlas}, and explain the experimentally observed hierarchy of encodings in Sec.~\ref{sec:hierarchy}.

\subsection{Full Santa Fe laser dataset simulation}
\label{sec:fullSFlas}

In order to showcase the performance of QRC in ideal conditions better, we performed simulations with the full Santa Fe laser dataset \cite{Weigend1993-bg} with 1400 train and 600 test data samples, including several regime switching events in both the train and test set. The parameters of the QRC and classical models were as reported in Sec.~\ref{sec:calcdetails}. In addition, the global pulse encoding was probed with both regular positions with $d=10 \, \mu \mathrm{m}$, and the irregular ones. The results are summarized in Tab.~\ref{tab:SFlas_perf}. The baseline performance level is provided by the naive model that predicts the next step in the timeseries to be the same as the last timestep in the window, and we take this performance to indicate no learning. Similarly to the MNIST data, we find that non-linear models, including 4-layer NN, position, and local pulse encoded QRC, all significantly outperform linear SVR and likely reach thresholds for the task at hand.

\begin{table}[htbp]
\centering
\begin{tabular}{@{}ccccc@{}}
\toprule
model &&   NMSE \\\midrule
naive ($t_{n+1}=t_n$)   && 0.96  \\
linear SVR   &&  0.21  \\
4-layer NN   && 0.0032  \\
 QRC (local)   && 0.004  \\
  QRC (position)   && 0.0038  \\
 QRC (pulse)  && 0.04  \\
  QRC (pulse, irregular positions)   && 0.025  \\ \bottomrule
\end{tabular}
\caption{\textbf{Simulated QRC and classical methods performance on the full Santa Fe laser timeseries task.} Lower NMSE is better. The classical models include the naive model (repeating the last point in the time window) and linear support vector regression (SVR), while three of the QRC encodings are simulated exactly. The pulse encoding is probed both with regular and fixed irregular atom positions. The quoted results are for 10-wide window features.
\label{tab:SFlas_perf}}
\end{table}

 The performance of the global pulse encoding, however, is between the linear SVR and the threshold non-linear performance. This performance lag can be explained by the effects of thermalization on the global pulse encoding, as discussed in Sec.~\ref{sec:hierarchy}. A partial improvement to the global pulse encoding performance is achieved by perturbing the atom positions from a regular to an irregular pattern (see also Sec.~\ref{sec:calcdetails}). We ascribe this performance improvement to the greater diversity in the encoding derived from early time-steps. Our simulations start from a translation-symmetric all-ground state, and in the regularly spaced chain case, this state evolves with the Hamiltonian with a weak translation symmetry breaking that is initially only felt close to the system edges. Thus, initial embeddings drawn respect this approximate translation symmetry and reduce the QRC expressiveness. Irregular atom positions strongly break translation symmetry, recovering the expressiveness of the early embeddings that are typical of other QRC encodings that all strongly break translation symmetry. Even with the irregular position improvement, the global pulse encoding lags behind thresholds reached with other encodings. 
 
\subsection{Hierarchy of encodings}
\label{sec:hierarchy}

In the simulations presented, a general picture of the approximate equivalence between local pulse and position encodings, and the weaker performance of global pulse encoding is seen. In exact simulation, we can explain the equivalence of position and local pulse encodings by the connection between position encoding and effective local detuning, as derived in Eq.~\ref{eq:encconnection}. The lag of the global pulse encoding in exact simulations stems from the fundamental limit induced by thermalization. Unique to the global pulse encoding, data features are encoded successively throughout the quantum dynamics. Due to thermalization \cite{Deutsch_2018}, the generic quantum dynamics at sufficiently long times results in the local observables independent of the exact details of the dynamics. This makes global pulse encoding of data lossy. The resulting encoding capacity limit does not apply to local pulse and position encodings where at least some of the early and intermediate-time (compared to thermalization time) drawn embeddings represent the whole data encoded. This limitation of the global pulse encoding could be circumvented by a process continuously probing information from the system, a possible example being protocols containing mid-circuit measurement and feedforward~\cite{Bravo2022}.

\begin{figure*}[htb]
\centering
\includegraphics[width=0.93\textwidth]{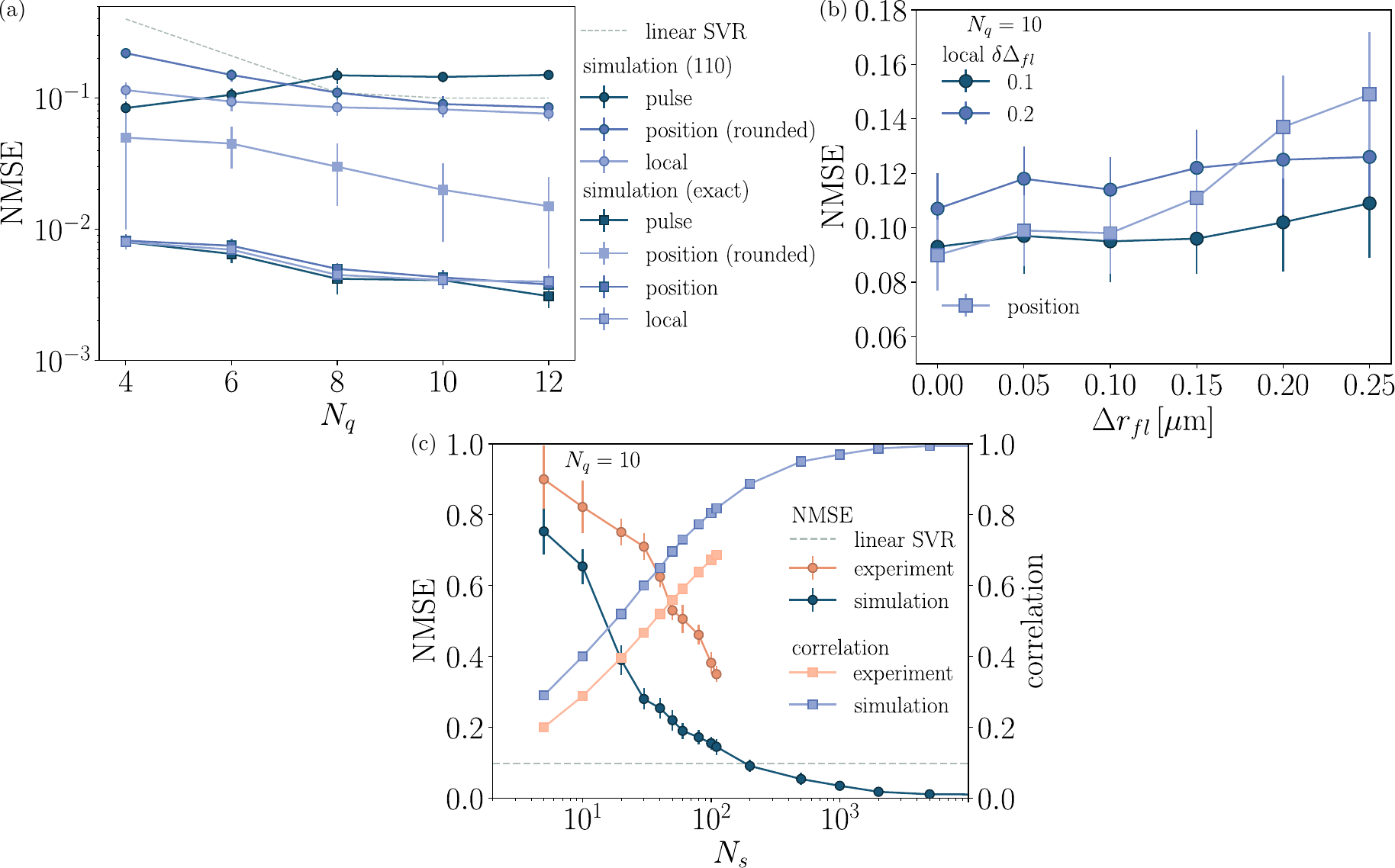}
\caption{\textbf{The details of numerically simulated and experimental performance of QRC on the timeseries prediction task.} (a) The NMSE (lower is better) dependence on the qubit number for numerically simulated QRC with all three encodings and the linear SVR for the Santa Fe laser task. The exactly simulated position encoding is shown both with precise position and the rounding of positions to $0.1 \, \mu$m precision, as required on the quantum hardware \cite{Wurtz2023}. (b) Numerically simulated QRC performance of the 10-qubit position and local pulse encoded approaches on the same task as a function of the RMS magnitude of shot-to-shot position fluctuations. The local pulse encoded results are shown for two different values of datapoint-to-datapoint local detuning fluctuations. The data shown is for 110 shots per datapoint drawn. (c) The NMSE (left axis, lower is better) and the training set correlation (right axis, higher is better) as the function of the number of shots drawn per datapoint ($N_s$) for 10-qubit global pulse encoded QRC in simulation and experiment. The linear SVR NMSE is shown for comparison. 
\label{fig:FigS4_SFlas}}
\end{figure*}

Santa Fe laser timeseries prediction task is particularly suitable for practical manifestations of the encoding hierarchy. Being regression, it allows performance differentiation that might be undetectable due to classification thresholds in the classification tasks. An example of encoding comparison in exact and finite Santa Fe laser task simulations is presented in Fig.~\ref{fig:FigS4_SFlas}(a), where the performance on the reduced dataset (the same as employed in experiments) is shown as a function of the qubit number for pulse parameters described in Sec.~\ref{sec:calcdetails}. In contrast to the full dataset, where global pulse encoding lagged behind, exact simulations show essentially equivalent performance for all three encodings, likely due to the absence of highly nonlinear regime switches in the test set. The performance consistently improves with increasing qubit number. The minute difference in the performance is readily seen; for example, the rounding of the atom positions to two decimal places directly affects encoding precision and performance. A similar but stronger effect is seen with finite sampling. With finite sampling and $N_s=110$ (corresponding to experiment), the ineffectiveness of global pulse encoding likely due to thermalization is again manifest, with system size scaling having the opposite trend from exact simulation. Local pulse and position encodings are essentially equivalent, with some differences due to the position rounding. Note that some of the differences between encodings stems from the changing window feature width in the case of the local pulse and position encodings, in contrast to the constant window width of 10 for the global pulse encoding. This emphasizes further the inverse qubit number scaling observed for the global pulse encoding.

We next turn to the question of the encoding hierarchy observed in experiments. There, global pulse encoding is the least effective, with the greatest discrepancy between noiseless simulation and experiment, followed by position and local pulse encodings. This necessitates the consideration of the experimental noise, where we generically expect two distinct scenarios:
\begin{itemize}
    \item Random noise that represents shot-to-shot fluctuations in QRC embeddings. The exemplary sources of the noise are the inherently stochastic nature of quantum sampling itself, shot-to-shot fluctuations in atom positions (for position encoding), and decoherence (for the position and local pulse encoding). The random noise does not present a significant obstacle to QRC performance, as the noise ensemble average mitigates its effects. In practice, the need for an ensemble average with various sources of random noise increases the effective sampling requirements of the QRC tasks.
    \item Systematic noise that arises when the two QRC embeddings drawn from the same data show statistically significant differences beyond what is expected due to finite sampling. A typical example is the calibration drift of the hardware that could result in inconsistent QRC embeddings on different pieces of the dataset. The systematic noise fundamentally affects the QRC performance; in practice, we attempt to manage it by statistical correlation tracking for different dataset batches, as described in Sec.~\ref{sec:consistency}. 
\end{itemize}

As an example of concrete effect, we first observe that the decoherence acts as a systematic noise for the global pulse encoding and the random noise for the other two encodings. This is due to the fact that global pulse encoding attempts to encode parts of the data in the already decohered quantum dynamics. The result is the lossy encoding for the same reason and compounding with the thermalization effects seen in the exact dynamics -- the two distinct data that only differ by later-encoded features could result in a similar embedding. This is likely the main driver behind the significant performance penalty for the experimental global pulse encoding compared to simulations.

In order to understand the performance hierarchy of the position and local pulse encoding observed in experiments, we quantify the effects of random noise stemming from the shot-to-shot uncertainty in the atomic positions present in the hardware \cite{Wurtz2023}. The corresponding simulations use the same parameters as the experimental protocols described in Sec.~\ref{sec:calcdetails}, with the addition of shot-to-shot fluctuations. The position fluctuations, implemented for both encodings, are generated by independent draws of the uniform random numbers in $[0, \Delta r_{fl}]$ interval for each shot, atom, and its two-dimensional cartesian coordinate. In addition, the random site-to-site fluctuations of the local detuning are also added to make a comparison closer to the quantum hardware due to their relevance as the source of systematic noise for the local pulse encoding. The local detuning fluctuations are implemented as shot-independent $\delta \Delta_{fl}$ relative fluctuations to the local detuning magnitude at each site. The main results of these simulations are presented in Fig.~\ref{fig:FigS4_SFlas}(b). While systematic site-to-site fluctuations affect the local pulse encoding, the effect of the shot-to-shot position fluctuations is very weak. This is in contrast with position encoding, which is more strongly affected by position fluctuations, particularly at larger position fluctuation amplitudes. While the noise model simulated here is by no means representative of the exact hardware noise model, the parameters $\Delta r_{fl}= 0.2 \, \mu \mathrm{m}$, $\delta \Delta_{fl}=0.1$ are likely the closest to the hardware observations~\cite{Wurtz2023}. Remarkably, at this noise level, the hierarchy between encodings and even the absolute performance level match well our experimental observations presented in Fig.~\ref{fig:Fig2_timeseries} of the main text.

Finally, we emphasize the practicality of the embedding statistical correlation as the proxy for noisy QRC performance. In Fig.~\ref{fig:FigS4_SFlas}(c), we present the shot scaling of the statistical correlation between the exactly simulated and finitely sampled simulated and experimental embeddings. The test set NMSE is shown for comparison. The statistical correlation was drawn from all the embeddings on the training set. It is clear from the data presented that both in the experiment and the finitely sampled simulation, the observed performance is inversely proportional to the embedding statistical correlation. Furthermore, the differences between the experimental and simulated correlation levels quantify the cumulative effects of all the noise sources and have a direct correspondence to the observed performance difference between experiments and finitely sampled simulation.

\begin{figure*}[htb]
\centering
\includegraphics[width=0.9\textwidth]{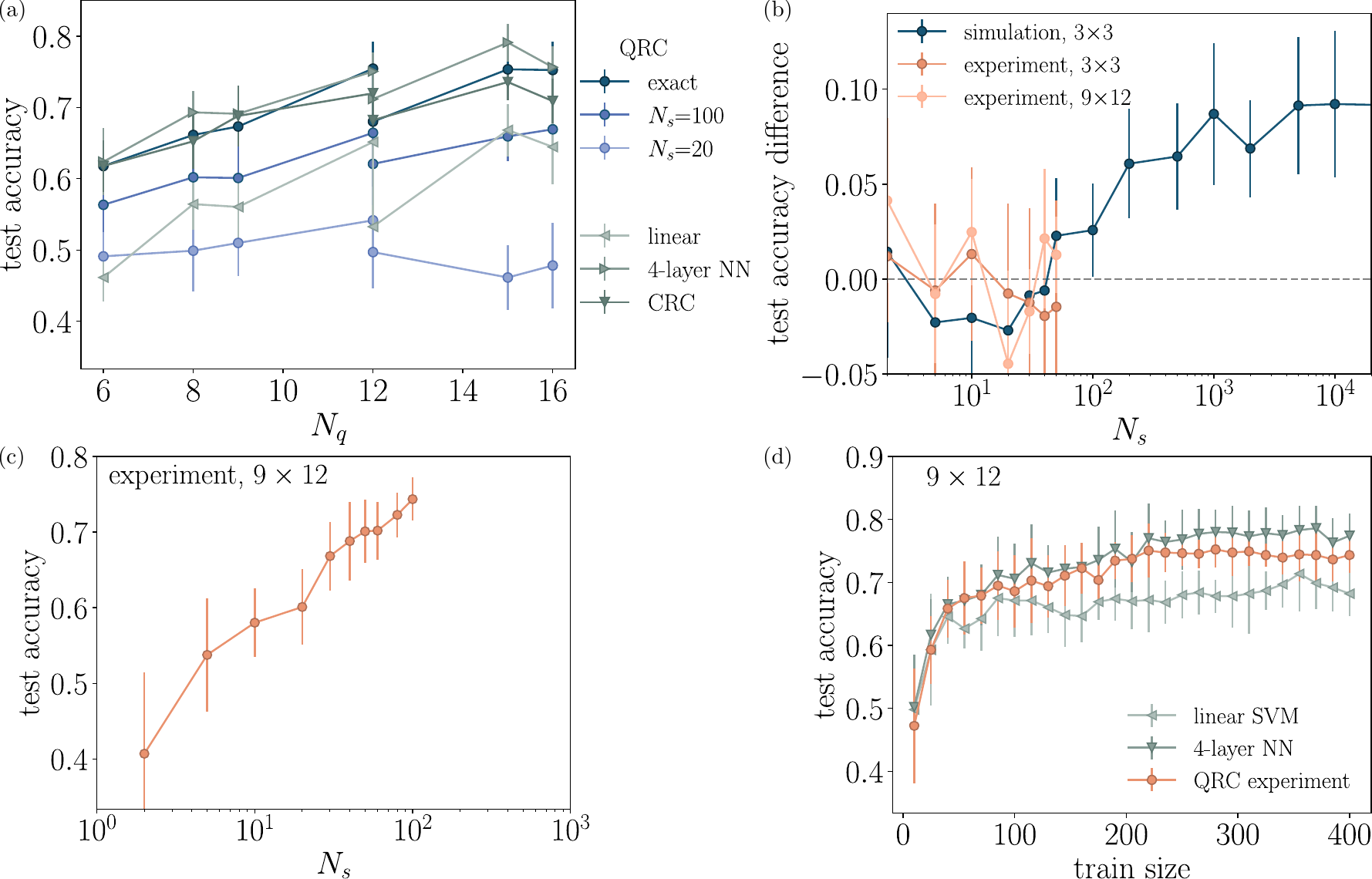}
\caption{\textbf{The details of numerically simulated and experimental performance of 2D QRC on the Tomato disease image classification task.} (a) Numerically simulated test accuracy of the position encoded QRC and several classical methods on the Tomato leaf disease dataset for several small system sizes. The QRC results are shown for several numbers of shots drawn per datapoint ($N_s$), including exact. (b) The numerically simulated and experimental QRC/CRC test accuracy difference on the kernel geometry constructed Tomato leaf task as a function of the number of shots drawn per datapoint. The system sizes of $3 \times 3$ and $9\times 12$ are shown. (c) The scaling of the test accuracy against the number of shots drawn per datapoint for the largest 108-qubit experimental task on the Tomato leaf dataset. (d) The dependence of the classification accuracy on the training data size (``training curve'') at $N_q=108$ for QRC experiment and classical methods. 
}
\label{fig:FigS5_Tomato}
\end{figure*}

\section{Supplementary results for multiclass image classification}
\label{sec:Tomato}

Here, we provide details of additional simulations and experimental results for the diseased tomato leaves classification task. In Fig.~\ref{fig:FigS5_Tomato}(a), the test accuracy of the simulated 2D QRC for several small 2D systems - 3$\times$2, 4$\times$2, 3$\times$3, 4$\times$3, 3$\times$4, 5$\times$3, and 4$\times$4 - and several $N_s$,  are shown. Similar to MNIST simulations, the QRC performance reaches classical nonlinear problem thresholds, with the finite $N_s$ providing the practical performance limit with a similar performance plateau as for MNIST. On the experimental side, the shot scaling for the largest system size employed is shown in Fig.~\ref{fig:FigS5_Tomato}(c). The scaling shows that the experiment is still outside of the plateau region for $N_s=100$, and thus direct performance gains are possible with increased sampling. A possible reason behind the successful system size scaling of QRC is the information redundancy that the large array and increased number of features provide. The embeddings in a larger system encode similar data features on different parts of the array, with their dimensions proportional to $N_q$. The total amount of bitwise shot data extracted from the quantum hardware is directly proportional to the system size as $N_s=100$ is kept constant.

The current experimental sampling limits the synthetic kernel data results presented in Fig.~\ref{fig:FigS5_Tomato}(b) as well. While the 3$\times$3 numerical simulations show significant comparative quantum kernel advantage, the breakeven point is around $N_s=100$. This is somewhat higher than the MNIST simulations, partially due to the smaller dataset size increasing the uncertainty of the results. As a result, the experimental results at the same system size that can only use $1/2$ of the shots for training and inference do not conclusively show CQKA. The sampling likely leads to the same inconclusive results on CQKA for the 9$\times$12 system.

Importantly, the QRC trainability is not adversely affected by the finite sampling or the increased embedding dimension stemming from the large system size. This is clear from Fig.~\ref{fig:FigS5_Tomato}(d), where the QRC performance as the function of the training set size (``training curve'') is similar to that of the two classical benchmarks even in the largest experiment performed.


\end{document}